\documentclass{article}

\usepackage[utf8]{inputenc}
\usepackage{subcaption}
\usepackage{graphicx}
\usepackage{amssymb}
\usepackage{xcolor}
\usepackage{hyperref}
\usepackage{tipa}

\usepackage{courier}

\usepackage{listings}
\definecolor{cppColorBackground}{rgb}{1.,1.,1.}
\definecolor{cppColorComment}{rgb}{0.0,0.47,.8}
\definecolor{cppColorLine}{rgb}{0.6,0.6,0.6}
\definecolor{cppColorString}{rgb}{0,0.501,145}
\definecolor{cppColorKey}{rgb}{0.8,0.5,0}
\definecolor{cppColorDigit}{rgb}{0,0,0.5}
\title{Specx: a C++ task-based runtime system for heterogeneous distributed architectures}

\author{
Paul Cardosi, B\'erenger Bramas\\
Inria Nancy, ICube Laboratory, University of Strasbourg, France\\
Berenger.Bramas@inria.fr
}

\begin{document}
\maketitle
\begin{abstract}
Parallelization is needed everywhere, from laptops and mobile phones to supercomputers.
Among parallel programming models, task-based programming has demonstrated a powerful potential and is widely used in high-performance scientific computing. 
Not only does it allow for efficient parallelization across distributed heterogeneous computing nodes, but it also allows for elegant source code structuring by describing hardware-independent algorithms.
In this paper, we present Specx, a task-based runtime system written in modern C++. 
Specx supports distributed heterogeneous computing by simultaneously exploiting CPUs and GPUs (CUDA/HIP) and incorporating communication into the task graph. 
We describe the specificities of Specx and demonstrate its potential by running parallel applications.
\end{abstract}



\section{Introduction}

Modern computers are increasingly heterogeneous and structured hierarchically, both in terms of memory and parallelization. 
This is especially visible in the high-performance computing (HPC) environment, where clusters of computing nodes equipped with multi-core CPUs and several GPUs are becoming the norm. 
Programming applications for this type of architectures is challenging, and using them efficiently requires expertise.

The research community has proposed various runtime systems to help parallelize computational codes. 
These tools differ on many aspects, including the hardware they target, their ease of use, their performance, and their level of abstraction. 
Some runtime systems have demonstrated flexibility in their use, but they are designed for experts, such as StarPU~\cite{augonnet:inria-00550877}. 
Others provide a modern C++ interface, but they do not support as many features as what HPC applications need, such as Taskflow~\cite{10.1109/TPDS.2021.3104255}.

In our current study, we describe Specx (\textipa{/'spEks/})~\footnote{\url{https://gitlab.inria.fr/bramas/specx}}, a runtime system that has been designed with the objective of providing the features of advanced HPC runtime systems, while being easy to use and allowing developers to obtain modular and easy-to-maintain applications.

The contribution of our work can be summarized as follows:
\begin{itemize}
\item We describe the internal organization of Specx, a task-based runtime system written in modern C++.
\item We present the key features needed to develop advanced HPC applications, such as scheduler customization, heterogeneous tasks, and dynamic worker teams.
\item We show that Specx allows developers to write compact C++ code thanks to advanced meta-programming.
\item Finally, we demonstrate the performance of Specx on several test cases.
\end{itemize}

The manuscript is organized as follows.
We provide the prerequisites in Section~\ref{sec:background} and the related work in Section~\ref{sec:relatedwork}.
Then, we describe Specx in Section~\ref{sec:specx}, before the performance study in Section~\ref{sec:study}.

\section{Background}
\label{sec:background}

In this section, we briefly describe task-based parallelization and the challenges it faces when computing on heterogeneous architectures.

\subsection{Task-based parallelization}
\label{sec:background:tb}

Task-based parallelization is a programming model in which the application is decomposed into a set of tasks. 
It relies on the principle that an algorithm can be decomposed in interdependent operations, where the output of some tasks is the input of others.
These tasks can be executed independently or in parallel, and they can be dynamically scheduled to different processing units while ensuring execution coherency.
The result can be seen as a direct acyclic graph (DAG) of tasks, or simple graph of tasks, where each node is a task and each edge is a dependency.
An execution of such a graph will start from the nodes that have no predecessor and continue inside the graph, ensuring that when a task starts, all its predecessors have completed.
The granularity of the tasks, that is, the content in terms of computation, cannot be too fine-grained because the internal management of the graph implies an overhead that must be negligible to ensure good performance~\cite{8314096}.
Therefore, it is usually the developer’s responsibility to decide what a task should represent.
The granularity is then a balance between the degree of parallelism and the runtime system overhead.
For that reason, several researches are conducted to delegate partially or totally the runtime system to the hardware with the objective of relieving the worker threads~\cite{10.1007/978-3-319-92040-5_20}.

The dependencies between tasks can be described in various ways. 
One way is to have the user explicitly connect tasks together. 
For example, the user might call a function $connect(t_i, t_j)$ to connect tasks $t_i$ and $t_j$. 
This approach requires the user to manage the coherency and to keep track of the dependencies between tasks, which can be error-prone and complicated between different stages of an application.
TaskFlow uses this approach.

Another way is to inform the runtime system about the input/output of the tasks, and letting it taking care of the coherency.
This approach is more convenient for the users, but there are many possibilities how this approach can be implemented.
One approach is to use a mechanism like the C++ future to access the result of asynchronous operations. 
This approach allows the runtime system to track the dependencies between tasks and ensure that they are satisfied without having a view on the input/output.
This approach is used by the ORWL runtime system~\cite{CLAUSS2010496}.

An alternative is to use sequential task-flow (STF)~\cite{Agullo:2016:IMS:2988256.2898348}, also called task-based data programming.
In this approach, the users describes the tasks and tells what are the data input/output for each task.
In general, a single thread creates the tasks and posts them in a runtime system, while informing about the access of each of them on the data.
The runtime system is then able to generate the graph and guarantee that the parallel execution will have the absolute same result as a sequential one.
This ends in a very compact code with few modifications required to add to an existing application by moving the complexity in the runtime system.
The sequential order is used to set the dependencies caused by read after write, or write after read.
This approach has a number of advantages, including:
\begin{itemize}
    \item A sequential program can be transformed into a parallel equivalent very easily.
    \item The users do not have to manage the dependencies.
    \item The accesses can be more precise than read/write and specific properties can be set to the accesses, such as commutativity.
    \item The tasks can be mapped to a graph, allowing the runtime system to analyze the graph to predict the workload or memory transfers and takes clever decisions.
\end{itemize}

In our work, we use the STF model.

\subsection{Computing on heterogeneous architectures}
\label{sec:background:hetero}

\begin{figure}
    \centering
    \includegraphics[width=.8\textwidth]{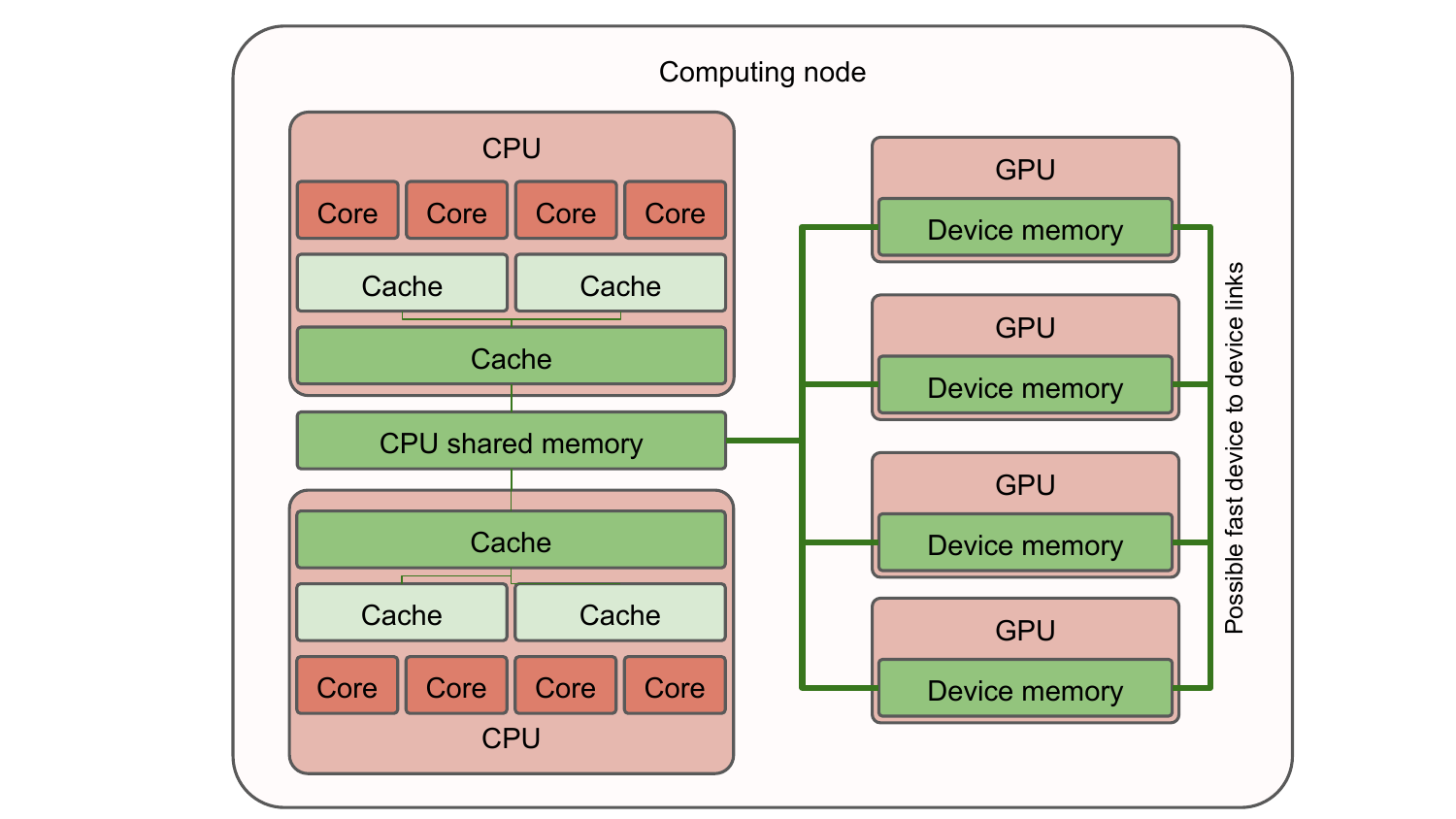}
    \caption{Simplified view of a heterogeneous computing node with 2 CPUs and 4 GPUs.
             Multiple nodes can be interconnected via network.}
    \label{fig:archi}
\end{figure}

Heterogeneous computing nodes consist of at least two distinct types of processing units. 
The most common configuration includes a dual-socket CPU paired with one or several GPUs, each having separate memory nodes (see Figure~\ref{fig:archi}). 
However, similar principles apply to other types of processing units as well.

Traditionally, these nodes operate in a pattern where a single CPU thread manages data movement to the device's memory, initiates the computational kernel, and waits for its completion before transferring the data back if required. 
To assist programmers, vendors have introduced unified memory, a mechanism that creates the illusion of a shared memory space between CPUs and GPUs. 
However, due to potential unpredictability and lack of control, its use remains rare in High-Performance Computing (HPC). 
Meanwhile, this usage pattern leaves other CPU cores idle, which is untenable in HPC.

To enhance the utilization of processing units, this pattern can be expanded to incorporate multiple CPU threads sharing a single GPU, enabled by mechanisms like streams or queues. 
This arrangement allows full exploitation of the GPU in terms of computational capability and memory transfers. 
However, managing the device's memory becomes increasingly complex, as it becomes crucial to avoid redundant object copying and ensure memory capacity isn't exceeded.

Furthermore, this method introduces the key challenge of balancing heterogeneous computing. 
Specifically, determining the optimal number of CPU threads per GPU, figuring out how to best use idle CPU cores, and deciding which parts of applications are more suited for CPUs than GPUs. 
In essence, how can we optimally distribute work among all processing units?

Task-based runtime systems aim to resolve these issues. 
They predominantly manage data transfers between CPUs and GPUs, allocate specific CPU cores to control GPUs, separate these cores from others to allow for concurrent executions, and schedule tasks across various types of processing units while considering workload and the most efficient processing unit type.

\section{Related work}
\label{sec:relatedwork}

\subsection{Task-based parallelization}

The most common task-based programming pattern can be described as a tasks-and-wait scheme, where independent tasks are inserted into a runtime system and a synchronization point allows waiting for their completion.
The task model from OpenMP 3~\cite{openmp08}\cite{ayguade2009design} and the task-based programming language Cilk~\cite{blumofe1996cilk} (later extended in Cilk++~\cite{leiserson2009cilkpp} and Cilk Plus~\footnote{https://www.cilkplus.org/}) follow this idea.
This remains a fork-join model because successive spawn phases of independent tasks (fork) must be explicitly synchronized (join) to ensure a correct execution.
Therefore, it limits the scalability because of the waiting time and the imbalance between tasks.
Of course, developers can increase the degree of parallelism by using multiple sources of tasks that they know are independent.
However, such an implementation starts to become a manual management of the dependencies, which a modern task-based runtime system is intended to do.

This is why there now exist numerous different task-based runtime systems that support dependency management.
The most popular ones are implementations of the OpenMP version 4~\cite{openmp13,chandra2001parallel} standard that defines the additional pragma keyword \emph{depend} to inform the runtime system about the type of data accesses performed by the tasks.
However, using pragmas, in general, is tedious when a task has hundreds of dependencies or when the number of dependencies are known at runtime. This can lead to ugly and error-prone code.
In addition, as OpenMP is a standard, it is upgraded slowly to ensure backward compatibility. 
Moreover, the standard is weak in the sense that it does not impose any constraints on the implementation and complexity of the underlying algorithms. 
This can cause performance surprises for the users when they use different OpenMP runtime systems.
In addition, OpenMP does not support distributed memory parallelization. 
Nonetheless, its portability, stability, and maturity make it a safe long-term choice.

StarPU~\cite{augonnet2011starpu} is a runtime system that was first designed to manage heterogeneous architectures. 
It is a C library, which means that users are constrained to use low-level programming and function pointers. 
However, it is extremely flexible and used by many HPC applications~\cite{agullo:inria-00547847,8671583,agullo2014task,lacoste2014taking,agullo2013multifrontal,9150424,8053812,moustafa2018task,bramas:hal-01942863}.
It also supports distributed memory parallelization with three different approaches~\cite{8226789}.
Each of these approaches uses a different description for the task graph, and the degree of information that the StarPU instances have on the complete graph is different. (there is one StarPU instance per computing node)
The first approach is the most trivial, and it consists of declaring the complete graph on all computing nodes. 
This means that there is one thread that describes the graph in each StarPU instance. 
Each instance can analyze the graph without any communication, since it holds the complete graph. 
This can be used to create low-cost scheduling strategies. 
For example, consider that all instances iterate over the task graph and have to decide on which computing node each task is executed. 
Thanks to the view on the complete graph, they can assign a task to a computing node while minimizing the communication, and all instances take the same decision without communicating. 
Moreover, the instances know where the data dependencies are located because they can track them while iterating on the graph. This can be used to post send/receive operations accordingly and manage the communication automatically.
However, there is a clear disadvantage to this approach: the method cannot scale because its cost and overhead increase with the size of the task graph, independently of the number of computing nodes that will be used.
In the second approach, each instance declares only a partial task graph that covers only the tasks it will compute. 
However, StarPU needs additional information to track the data movement and to connect the different partial task graphs that manage the communication. 
The first option consists in requesting explicit communication calls (similar to the MPI) that connect the tasks between the instances. 
In the second option, each instance inserts the tasks that will be computed by others and which are at the frontier of its partial graph.
These two approaches remove abstraction because the developers manually split the task graph and have to manage the boundaries of the partial graph. Moreover, each instance has only a partial view, making analysis and scheduling difficult. 
Specx uses a similar approach.

PaRSEC~\cite{bosilca2013parsec,danalis2014ptg} is a runtime system based on the parametrized task graph (PTG) model~\cite{375471}.
It has been demonstrated to be effective in various scientific applications.
The PTG is a domain-specific language (DSL) that captures a static, algebraic description of a task graph that can be expanded efficiently at runtime. 
This allows PaRSEC to manage large graphs without fully instantiating them. 
This approach works well on affine loops thanks to polyhedral analysis.
The analysis of the data-flow of a task instance is constant in time, and the representation of the graph is constant in space. 
This makes the PTG a very efficient way to represent task graphs.
However, the PTG is not as expressive as other task graph models. 
It is difficult to use the PTG to represent applications with irregular or sparse algorithmic or data access patterns.
Despite this limitation, the PTG has been shown to be effective in a wide variety of scientific applications. 
It is a powerful tool for parallelizing applications that can be expressed in terms of affine loops.
It is theoretically possible to write PTGs for highly dynamic applications, but this would imply an unbounded amount of time building and traversing dynamic meta-data in memory.
However, the PTG is impractical for implementing applications with irregular or sparse algorithmic or data access patterns, where the logic is difficult to express with linear equations.
The PTG graph representation is highly efficient, but the expressiveness of the model is limited. 
Internally, the representation allows collapsing a task graph in two dimensions, i.e., time and parallelism~\cite{soi2021index}, which permits several optimizations.
In distributed-memory, the different Parsec instances all hold an algebraic representation of the complete graph. 
Parsec uses advanced mechanisms to allow scheduling the tasks efficiently using heuristics and potential input from the users.

Charm++~\cite{kale1993charmpp,10.1145/167962.165874,chamberlain2007parallel} is an object-oriented parallel programming framework that relies on a partitioned global address space (PGAS) and that supports the concept of graphs of actors. 
It includes the parallelism by design with a migratable-objects programming model, and it supports task-based execution.
The actors (called \emph{chares}) interact with each other using invocations of asynchronous methods. 
However, with Charm++, there is no notion of tasks as we aim to use. 
Instead, tasks are objects that communicate by exchanging messages.
Charm++ schedules the chares on processors and provides object migration and a load balancing mechanism. 
PGAS allows accessing data independent of their actual location, which is the inverse of what the task-based method intends to offer. 
A task is a piece of work that should not include any logic or communication. 
This approach forbids many optimizations and mechanisms that task graphs support~\cite{thibault:tel-01959127}.

HPX~\cite{kaiser2014hpx} is an open-source implementation of the ParalleX execution model. 
Its implementation aims to respect the C++ standard, which is an asset for portability and compliance with existing C++ source code.
In HPX, tasks request access to data by calling an accessor function (get/wait). 
The threads provide the parallelism description, which is tied to the order and type of data accesses. 

OmpSs~\cite{duran2011ompss,perez2008dependency} uses the insert-task programming model with pragmas similar to OpenMP through the Nanos++ runtime to manage tasks. 
When running in distributed memory, it follows a master-slave model, which may suffer from scalability issues as the number of available resources or the problem size increase.

XKaapi~\cite{gautier2013xkaapi} is a runtime system that can be used with standard C++ or with specific annotations, but it requires a specific compiler.
Legion~\cite{bauer2012legion} is a data-centric programming language that allows for parallelization with a task-based approach.
SuperGlue~\cite{tillenius2015superglue} is a lightweight C++ task-based runtime system. It manages the dependencies between tasks using a data version pattern.
X10~\cite{charles2005x10} is a programming model and a language that relies on PGAS, too, and hence has similar properties as Charm++.
Intel Threading Building Blocks~\footnote{https://www.threadingbuildingblocks.org/} (ITBB) is an industrial runtime system provided as a C++ library. 
It is designed for multicore parallelization or in conjunction with oneAPI, but it follows a fork-join parallelization pattern.

Regarding distributed parallelization, most runtime systems can be used with MPI~\cite{snir1998mpi}.
The developers implement a code that alternate between calls to the runtime systems and post of MPI communications.
When supported by the runtime system, the data movements between CPUs/GPUs and in-node load balancing are delegated to the runtime system.
More advanced methods have been elaborated, and they entirely delegate the communications to the runtime system~\cite{ZAFARI2019102582,10.1007/978-3-319-78024-5_16,10.1007/978-3-319-98521-3_15,8622893,8565930}, like Parsec, StarPU~\cite{agullo:hal-01332774}, Legion, Charm++, TaskTorrent~\cite{cambier2020tasktorrent}, and HPX.

Most of these tools support a core aspect of a task-based runtime system, including the creation of a task graph (although the implementation may vary) where tasks can read or write data.
However, scheduling is an important factor in the performance~\cite{agullo2016task}, and few of these runtime systems propose a way to create a scheduler easily without having to modify the code.
Moreover, specific features offer mechanisms to increase the degree of parallelism.
For instance, some runtime systems permit the specification of whether data access is commutative, implying that tasks write data without any particular order.
This kind of advanced functions can significantly impact performance~\cite{agullo2017bridging}.
They differ whether the task graphs are statically or dynamically generated, how the generation is performed, which in-memory representation is used, which parallelization levels are supported, and many other features~\cite{thomantaxonomy, thoman2018taxonomy,gu2019comparative,gurhem2020current}.


\subsection{Speculative execution}

Speculative execution is an approach that can increase the degree of parallelism.
It has been widely used on hardware and is an ongoing research topic in regard to software~\cite{Survey16,8114234,doi:10.1002/cpe.4192}.
The key idea of speculative execution is to utilize idle components to execute operations in advance, which includes the risk of performing actions that may later be invalidated.
The prominent approach is to parallelize an application, and to detect at runtime if race conditions or accesses with invalid orders which violate dependencies happen.
The detection of invalid speculative execution can be expensive, and as a result, some research is intended to design hardware modules for assistance~\cite{10.1145/342001.339650,8038067}.
However, these low-level strategies are unsuitable for massively parallelized applications and impose the need for either detecting the code parts suitable for speculation or relying on explicit assistance from developers.

In a previous study, we have shown that characterizing accesses as 'maybe-write' instead of 'write' allows us to increase the degree of parallelism thanks to speculative execution in the task-based paradigm~\cite{10.7717/peerj-cs.183}.
This novel kind of uncertain data access (UDA) can be used when it is uncertain at task insertion time whether the tasks will modify some data or not.
Similar to the 'commutative write', developers simply provide additional information to the runtime system, enabling it to set up a strategy by modifying the graph of tasks on the fly.
This also makes it possible to delay some decisions from the implementation time to the execution time, where valuable information about the ongoing execution is available.
We have implemented this mechanism in our task-based runtime system Specx (originally called SPETABARU) and conducted an evaluation on Monte Carlo simulations, which demonstrated significant speedups.
We are currently developing a new model~\cite{souris:hal-04156383}.

On different fields, speculation has also been used in a tasking framework for adaptive speculative parallel mesh generation~\cite{tsolakis2022tasking} and for resource allocation in parallel trajectory splicing~\cite{garmon2022resource}.

\section{Specx's features, design and implementation}
\label{sec:specx}

\subsection{Task graph description}

In Specx, we dissociated the task-graph from the so-called computing engine that contains the workers.
Therefore, the user has to instantiate a task-graph and select among two types, one with speculative execution capability and one without, which allows the removal of the overhead of the speculative execution management when no UDAs will be used.
We provide an example in Code~\ref{code:runtime}.

\begin{figure}[h!]
\begin{lstlisting}[escapechar=|, label={code:runtime}, caption={Specx example - creation of a task graph.}]
// Create a task graph
SpTaskGraph<SpSpeculativeModel::SP_NO_SPEC> tg;
// Legacy version, create a runtime (a compute engine + a task graph)
SpRuntime runtime(SpUtils::DefaultNumThreads());
\end{lstlisting}
\end{figure}

\paragraph{Task Insertion}
Specx follows the STF model: a single thread inserts the task in the runtime system (task-graph object) and tells which variables will be written or read.
Additionally, the user can pass a priority that the scheduler is free to use when making decisions.
The core part of the task consists of a callable object with the operator \emph{()}, which allows for the use of C++ lambda functions.
The data access modes that Specx currently supports are:
\begin{itemize}
    \item SpRead: the given dependency will only be read by the task.
 As such, the parameter given to the task function must be \emph{const}.

    \item SpWrite: the given dependency will be read or write by the task.
    
    \item SpCommutativeWrite: the given dependency will be read or write by the task but the order of execution of all the \emph{SpCommutativeWrite} inserted jointly is not important.

    \item SpMaybeWrite: the given dependency might be read and/or write by the task.
 Possible speculative execution patterns can be applied.

    \item SpAtomicWrite: the given dependency will be read or write by the task, but the user will protect the access by its own mechanisms (using mutual exclusion, for example).
 The runtime system manages this access very similarly to a read access (multiple \emph{SpAtomicWrite} can be done concurrently, but the runtime system has to take care of the read-after-write, write-after-read coherency).
\end{itemize}

When a dependency \emph{X} is passed, the runtime dereferences \emph{X} to get its address, and this is what will be used as the dependency.
An important point when using task-based programming is that it is the user's responsibility to ensure that the objects will not be destroyed before all tasks that use them are completed.
We provide an example in Code~\ref{code:task}.

\begin{figure}[h!]
\begin{lstlisting}[escapechar=|, label={code:task}, caption={Specx example - creation of a task for CPU.}]
const int initVal = 1;
int writeVal = 0;
// Create a task with lambda function
tg.task(SpRead(initVal), SpWrite(writeVal),
             [](const int& initValParam, int& writeValParam){
    writeValParam += initValParam;
});
\end{lstlisting}
\end{figure}

\paragraph{Dependencies on a Subset of Objects}

A critical drawback of OpenMP 4 was the rigidity of the dependency declaration. 
Indeed, the number of dependencies of a task had to be set at compile time.
Since OpenMP 5, it is now possible to declare an iterator to express dependencies on several addresses at once.
This is extremely useful if, for example, we want to declare a dependency on all the elements of a vector.

To solve this issue, in Specx, we can declare the dependencies on a set of objects using the following mechanisms: \emph{SpReadArray($<$XTy$>$ x, $<$ViewTy$>$ view)}, \emph{SpWriteArray($<$XTy$>$ x, $<$ViewTy$>$ view)}, \emph{SpMaybeWriteArray($<$XTy$>$ x, $<$ViewTy$>$ view)}, \emph{SpCommutativeWriteArray($<$XTy$>$ x, $<$ViewTy$>$ view)}, \emph{SpAtomicWriteArray($<$XTy$>$ x, $<$ViewTy$>$ view)}, where \emph{x} should be a pointer to a contiguous buffer (or any container that support the \emph{[]} operator), and \emph{view} should be an object representing the collection of specific indices of the container elements that are affected by the dependency. 
\emph{view} should be iterable (in the sense of "stl iterable").

With this mechanism, Specx can iterate over the elements and apply the dependencies on the selected ones.
We provide an example in Code~\ref{code:arrayview}.

\begin{figure}[h!]
\begin{lstlisting}[escapechar=|, label={code:arrayview}, caption={Specx example - use array of dependencies.}]
std::vector<int> vec = ...;
// Access all the elements in the SpArrayView
tg.task(SpPriority(1), SpWriteArray(vec.data(),SpArrayView(vec.size())),
            [](SpArrayAccessor<int>& vecView){
    ...
});
\end{lstlisting}
\end{figure}

\paragraph{Task Viewer}

Inserting a task in the task-graph returns a task view object, which allows accessing some attributes of the real task object. 
For instance, it allows setting the name of the task, waiting for the task completion, or getting the value produced by the task (in case the task returns a value).
Unfortunately, there is a pitfall with the current design, which is the fact that accessing the task through the viewer can potentially be done after the task has been computed. 
For instance, we cannot use the tasks' names in the scheduler because they might be set after the tasks are computed.
We provide an example in Code~\ref{code:taskviewer}.

\begin{figure}[h!]
\begin{lstlisting}[escapechar=|, label={code:taskviewer}, caption={Specx example - task viewer.}]
auto taskViewer = runtime.task(SpRead(initVal), SpWrite(writeVal),
             [](const int& initValParam, int& writeValParam) -> bool {
    writeValParam += initValParam;
    return true;
});
taskViewer.setName("The name of the task");
taskViewer.wait(); // Wait for this single task
taskViewer.getValue(); // Get the value (when the task is over)
\end{lstlisting}
\end{figure}

\subsection{Teams of Workers and Compute Engines}

Within Specx, a team of workers constitutes a collection of workers that can be assigned to computational engines. 
In the current implementation, each worker is associated with a CPU thread that continuously retrieves tasks from the scheduler and handles them. 
If the worker is CPU-based, the task is directly executed by the CPU thread. 
Conversely, in the case of a GPU worker, the CPU thread manages the data movement between memory nodes and call the device kernel.

A compute engine necessitates a team of workers and may be responsible for several task-graphs. 
Currently, it is not possible to change the compute engine assigned to a task-graph, but it is possible to shift workers among different compute engines. 
This feature provides the ability to dynamically adjust the capabilities of the compute engine during execution and design advanced strategies to adapt to the workload of the graphs.

Given that dependencies among task-graphs are not shared, the insertion of tasks and their dependencies into a task-graph does not affect others. 
This allows for the creation of recursive parallelism, in which a task-graph is created within a task. 
Such a task-graph could potentially be attached to the same compute engine as the parent task. 
This approach could help mitigate the overhead associated with the creation of a large set of tasks by organizing them into sub-task-graphs.
We provide an example in Code~\ref{code:teams}.

\begin{figure}[h!]
\begin{lstlisting}[escapechar=|, label={code:teams}, caption={Specx example - creation of a compute engine.}]
SpTaskGraph<SpSpeculativeModel::SP_NO_SPEC> tg;
// Create the compute engine
SpComputeEngine ce(SpWorkerTeamBuilder::TeamOfCpuWorkers(NbThreads));
// OR
SpComputeEngine ce(SpWorkerTeamBuilder::TeamOfCpuCudaWorkers());
// Tells which compute engine will manage the graph
tg.computeOn(ce);
\end{lstlisting}
\end{figure}

\subsection{Tasks for Heterogeneous Hardware}

Specx relies on the same principles as StarPU in supporting heterogeneous hardware, i.e., we have distinct workers for each type of processing unit, and each task can operate on CPUs, GPUs, or both. 
Specifically, at task insertion, we require a unique callable object for each processing unit type capable of executing the task. 
During execution, the scheduler determines where the task will be executed.
This represents a critical challenge in task-based computing on heterogeneous systems.

Regarding the interface, the primary challenge is the movement of data between memory nodes. 
More specifically, we strive to exploit C++ and use an abstraction mechanism to facilitate object movement. 
Consequently, we have determined that objects passed to tasks should comply to one of the following rules:
1) the object is trivially copyable~\footnote{\url{https://en.cppreference.com/w/cpp/types/is_trivially_copyable}};
2) the object is a std::vector of trivially copyable objects;
3) the object's class implements specific methods that the runtime system will call.

In the last case, the object's class must have as a class attribute a data type called \emph{DataDescriptor} and three methods:
\begin{itemize}
    \item \emph{memmovNeededSize}: Invoking this method on the object should yield the required size of the memory to be allocated on the device for copying the object.
    \item \emph{memmovHostToDevice}: This method is called to transfer the object to the device. The method receives a mover class (with a copy-to-device method) and the address of a memory block of the size determined by \emph{memmovNeededSize} as parameters. The method may return a \emph{DataDescriptor} object, which will later be passed to \emph{memmovDeviceToHost} and to the task utilizing the object.
    \item \emph{memmovDeviceToHost}: This method is invoked to move the data back from the GPU to the object. The method receives a mover class (with a copy-from-device method), the address of a GPU memory block, and an optional \emph{DataDescriptor} object as parameters.
\end{itemize}

From a programming perspective, we require the users to determine how the data should be moved as they have the knowledge to do so. 
For example, consider an object on a CPU being a binary tree where each node is a separate memory block.
It would be inefficient to allocate and copy each node. 
Consequently, we ask the users to estimate the needed memory block size, and we perform a single allocation. 
Then it is the users' responsibility to mirror the tree on the GPU using the block we allocated, and to implement the task such that it can use this mirror version. 
This design may change in the future as we continue to apply Specx to existing applications.

Currently, we employ the Least Recently Used (LRU) policy to determine which memory blocks should be evicted from the devices when they are full.
Concretely, this implies that when a task is about to be computed on the device, the worker's thread will iterate over the dependencies and copy them onto the GPU's memory using a stream/queue. 
If an object already has an up-to-date version on the device, the copy will be skipped, and if there is not enough free memory, older blocks may be evicted. 
As a result, at the end of a simulation, the up-to-date versions of the objects might be on the GPUs, necessitating their transfer back to the CPUs if required. 
At present, this can be accomplished by inserting empty CPU tasks that use these objects.

By default, worker teams align with hardware configurations, i.e., they will contain GPU workers for each available GPU. Therefore, if users are not careful and only need one type of processing unit for their tasks, the hardware will be underutilized as some workers will remain idle.

We provide an example of task for both CPU and GPU, and a class that provides a copy method in Code~\ref{code:cuda}.

\begin{figure}[h!]
\begin{lstlisting}[escapechar=|, label={code:cuda}, caption={Specx example - creation of a task for CPU/GPU.}]
class Matrix{
    int nbRows;
    int nbCols;
    std::vector<double> values;
public:
    // What to allocate on the device
    std::size_t memmovNeededSize() const{
        return sizeof(double)*nbRows*nbCols;
    }

    // Copy to the device (size == memmovNeededSize())
    template <class DeviceMemmov>
    auto memmovHostToDevice(DeviceMemmov& mover, void* devicePtr, std::size_t size){
        double* doubleDevicePtr = reinterpret_cast<double*>(devicePtr);
        mover.copyHostToDevice(doubleDevicePtr, values.data(), nbRows*nbCols*sizeof(double));
        return DataDescr{rowOffset, colOffset, nbRows, nbCols};
    }

    // Copy to the CPU
    template <class DeviceMemmov>
    void memmovDeviceToHost(DeviceMemmov& mover, void* devicePtr, std::size_t size, const DataDescr& /*inDataDescr*/){
        double* doubleDevicePtr = reinterpret_cast<double*>(devicePtr);
        mover.copyDeviceToHost(values.data(), doubleDevicePtr,  nbRows*nbCols*sizeof(double));
    }
};

// ....
Matrix matrix;

tg.task(SpPriority(1), SpWrite(matrix),
    SpCpu([](Matrix& matrix){
        // ...
    })
#ifdef SPECX_COMPILE_WITH_CUDA
  , SpCuda([](SpDeviceDataView<Matrix> matrix) {
        // ...
    })
#endif
).setTaskName(std::string("My operation"); // Set the name of the task
\end{lstlisting}
\end{figure}

In most GPU programming models, a CPU thread must explicitly request the use of one of the GPUs available on a machine (e.g., using \texttt{cudaSetDevice}).
Once a device is selected, all subsequent calls to the GPU will be executed on that specific device.
However, selecting a device incurs overhead due to the creation of a context.
When a CPU thread needs to compute a task that requires updated data stored in the memory of one or several GPUs, it must first copy the data to the system's RAM.
To address this, we have implemented a background CPU thread for each GPU, which can interact with the GPU.
When a CPU worker needs to interact with a GPU, it does so through the background thread using a spawn/sync mechanism.

\subsection{Mixing Communication and Tasks}

In the context of distributed memory parallelization, Specx provides the capability to mix send/receive operations (MPI) and computational tasks. 
Putting MPI communications directly inside tasks will fail due to the potential concurrent accesses to the communication library (which is not universally supported by MPI libraries) and the risk of having workers waiting inside tasks for communication completion, leading to deadlocks if tasks sending data on one node do not coincide with tasks receiving data on another. 
Therefore, to avoid having the workers dealing with communication, our solution is to use a dedicated background thread that manages all the MPI calls.

In this approach, a \emph{send} operation is transformed into a communication task that does a write access on the data, with execution will be done by the background thread.
Similarly, a \emph{receive} operation becomes a communication task that performs a read access, and is also managed by the background thread. 
Once a communication task is ready, the background thread executes the corresponding non-blocking MPI calls, receiving an MPI request in return. 
This request is stored in a list, which the background thread aims to complete by calling the MPI \emph{test-any} function.
When a request is fulfilled, the background thread releases the dependencies of the associated communication task, thereby ensuring progression of the task-graph execution.
In this way, the progression is done as early as possible.

In order to send/receive C++ objects using MPI in a single communication (although we perform two - one for the size and one for the data), we need a way to store the object into a single array. 
To achieve this, the object must comply with one of the following rules:
\begin{itemize}
    \item It should be trivially copyable;
    \item It should provide access to a pointer of the array to be sent (or received). For example, if a class has virtual methods, it will not be trivially copyable. However, if the class's only attribute is a vector of integers, sending the object is equivalent to sending the vector's data.
    \item It should support our serialization/deserialization methods. Here, we allow the object to serialize itself using our utility serializer class, yielding a single array suitable for communication. 
    Upon receipt, the buffer can be deserialized to recreate the object.
    This method offers the more flexibility, but is also the less efficient.
\end{itemize}

Specx also supports MPI broadcast as part of MPI global communication functions. Currently, users must ensure that all Specx instances perform the same broadcasts and in the same order.

As a side note, MPI communications are incompatible with the speculative execution capabilities of Specx due to the potential creation of extra tasks and instantiation of diverse execution paths.

We provide an example of distributed implementation in Code~\ref{code:mpi}.

\begin{figure}[h!]
\begin{lstlisting}[escapechar=|, label={code:mpi}, caption={Specx example - sending/receiving a matrix object (using the serializer method).}]
class Matrix{
    // ...
    Matrix(SpDeserializer &deserializer)
        : nbRows(deserializer.restore<decltype(nbRows)>("nbRows")),
          nbCols(deserializer.restore<decltype(nbCols)>("nbCols")),
          values(deserializer.restore<decltype(values)>("values")){
    }

    void serialize(SpSerializer &serializer) const {
        serializer.append(nbRows, "nbRows");
        serializer.append(nbCols, "nbCols");
        serializer.append(values, "values");
    }
};

// ....
tg.mpiRecv(matrix, srcRank, tag);
// ....
tg.mpiSend(matrix, destRank, tag);
\end{lstlisting}
\end{figure}

\subsection{Scheduling}

We designed the scheduler module following the implementation approach used in StarPU, with the scheduler providing two key functions: \emph{push} and \emph{pop}. When a task becomes ready (i.e., its predecessors have finished), it is pushed into the scheduler. Conversely, when a worker becomes available, it calls the pop function on the scheduler, which may potentially return \texttt{null} if there are no tasks compatible with its processing unit type, or if the scheduler determines it is not appropriate to assign a task. Thus, the scheduler plays a crucial role in managing task distribution and the order of task execution.

At present, Specx includes several simple schedulers. The first one is a First-In-First-Out (FIFO) scheduler, called \texttt{SpSimpleScheduler}, which uses a single list implemented with atomic operations. This scheduler only supports CPU workers.

The second scheduler, \texttt{SpPrioScheduler}, uses a priority queue to store tasks, ensuring they are popped in priority order. However, this scheduler also only supports CPU workers.

The third scheduler, called \texttt{SpHeterogeneousPrioScheduler}, uses three priority queues: one for CPU-only tasks, one for GPU-only tasks, and one for hybrid tasks. Workers first pick a task from their corresponding queue and then from the hybrid queue if necessary. Consequently, tasks within a queue are consumed in priority order, but tasks of higher priority might reside in the hybrid queue.

Finally, we provide a variant of the heteroprio scheduler~\cite{agullo2016task,10.7717/peerj-cs.969}. In our implementation, each GPU worker has its own queue. The user should provide a priority-pair with each task, where the pair represents a priority for the CPU and a priority for the GPU. Thus, tasks are inserted into the queue with different priorities for the CPU and the GPU (the idea being that a task for which the GPU is faster should have a higher priority on the GPU, and vice versa for the CPU). This scheduler includes an option to favor the worker that releases the task by providing a bonus to the task’s priorities.

We plan to introduce more sophisticated schedulers~\cite{10596474} in the near future, but this will require performance models of the tasks to enable the scheduler to make more informed decisions, which implies significant engineering work.

In any case, with our design, it is straightforward to provide a new scheduler, and users have the flexibility to implement a custom scheduler specifically designed for their application. This can be accomplished by creating a new class that inherits from our abstract scheduler interface.

\subsection{Speculative execution}

Specx supports task-based speculative execution~\cite{10.7717/peerj-cs.183}, which is an ongoing research problem. 
We currently support two speculative models applicable when certain data accesses are flagged as \emph{maybe-write}. 
In these scenarios, the runtime system may duplicate some data objects and tasks to enable potential speculative work, subsequently performing a rollback if the uncertain tasks actually modified the data.

\subsection{Internal implementation}

In this section, we delve into the finer details of Specx's implementation. 
When a task is inserted, the callable's prototype should match with the dependency types.
Hence, read parameters are passed as \emph{const}. 
For CPU callables, parameters should be of the type references to the object types passed as arguments. 
Using a value instead of a reference will simply result in a copy, which is typically not the intended outcome. 
Indeed, if values are required but are not significant as dependencies, it is more appropriate to pass them as captures in the lambda.

When an object is passed to a task, the runtime system dereferences it to obtain its address. 
This address is utilized as a dependency value and also as a key in an unordered hashmap that matches pointers to data handles.
A data handle is a class that contains all the necessary information the runtime system requires concerning a dependency. 
For instance, the data handle includes the list of dependencies applied to the associated object.
This allows progression in the list and subsequent release of dependencies.
In terms of implementation, we do not construct a graph; instead, we have one data handle per address that has been used as a dependency, and the data handles' dependency lists contain pointers to the tasks that use the related objects.
Consequently, when a task is finished, we increment a counter on the dependency list and access the next tasks. 
Doing so, we then examine whether the task now pointed to by the updated counter is ready, and we push the task into the scheduler if that is the case.
The data handle also possesses a mutual exclusion object that enables its locking for modification. 
When several data handles need to be locked, we sort them based on their address, ensuring deadlock prevention.

The commutative (\emph{SpCommutativeWrite}) dependency is managed differently because the related tasks' order is not static.
Said differently, when the next tasks use the commutative access, we do not know which one should be executed.
As such, we cannot merely point to the first task in the dependency list and stop our inspection if it's not ready, as the following tasks that also have commutative access to the dependency might be ready. 
This necessitates a check on all the tasks performing a commutative access at the same time point. 
However, several threads that completed a task might do so simultaneously, requiring us to use a mutual exclusion to protect all the commutative dependencies. 
In other words, using commutative dependencies implies the use of global mutual exclusion.

We use C++ meta-programming massively, such as testing if an object is trivially copyable or supports serialization methods, for example.
We also utilize the inheritance/interface pattern and the template method design pattern. For instance, this enables us to have a task class containing the callable type, and thus the types of all parameters and arguments.
We can then carry out meta-programming tests on the arguments to ensure compliance with specific rules, etc.

Finally, as we use a hashmap to store the data dependency objects' information, using their address as keys, it's currently undefined behavior to have objects of different types but stored at the same address.
This primarily occurs when an object of type $x$ is freed, and subsequently, an object of type $y$ is allocated using the same memory block.
This is because the data handle class uses a data copier through an interface, but the copier is actually templatized over the dependency object type.

\subsection{Visualization}

Profiling and optimizing task-based applications are crucial to achieve high performance. The main information are:
\begin{itemize}
\item Degree of parallelism: This represents how many tasks can be executed in parallel.
The task graph can be used to evaluate if the degree of parallelism is sufficient to utilize all the processing units fully. Furthermore, during the execution, the number of ready tasks over time can also be analyzed.
\item Task granularity: The task granularity can impact the degree of parallelism. 
An examination of an execution trace can help determine if the granularity is too small.
If so, the overhead of task management and/or data displacement may be too large compared to the task durations, thereby negatively affecting performance.
Conversely, if the tasks are too large, the degree of parallelism can be too small, and the end of the execution can be penalized with too few large tasks to compute.
\item Scheduling choices for task distribution: If a slow worker is selected mistakenly (a worker that can compute a task but is not efficient to do so), it can reduce performance. It could be faster to wait and assign the tasks to a quicker worker, but this depends on the scheduler.
\item Scheduling choices for task order: The degree of parallelism (and sometimes the availability of suitable tasks for all workers) can be influenced by the order of task execution, that is, the choice among the ready tasks.
\end{itemize}

A good situation, but not necessarily optimal, is when no workers have been idle, and the tasks have been assigned to the processing units that can execute them most efficiently.

To facilitate the profiling, Specx provides features to export the task-graph and execution trace.
The task-graph is generated in the \emph{dot} format.\footnote{\url{https://gitlab.com/graphviz/graphviz}}
For the execution trace, a SVG file\footnote{\url{https://www.w3.org/Graphics/SVG/}} is exported that can be opened with any modern internet browser.
The execution trace also indicates the number of tasks available during the execution.
Specx also allows the export of the amount of memory transferred between the CPUs and the GPUs.
In the next release, we plan to export metrics that will provide concise but meaningful numbers on execution quality, such as the idle time.

\begin{figure}[h]
    \centering        
    \begin{subfigure}{0.47\textwidth}
        \centering
        \includegraphics[width=\textwidth]{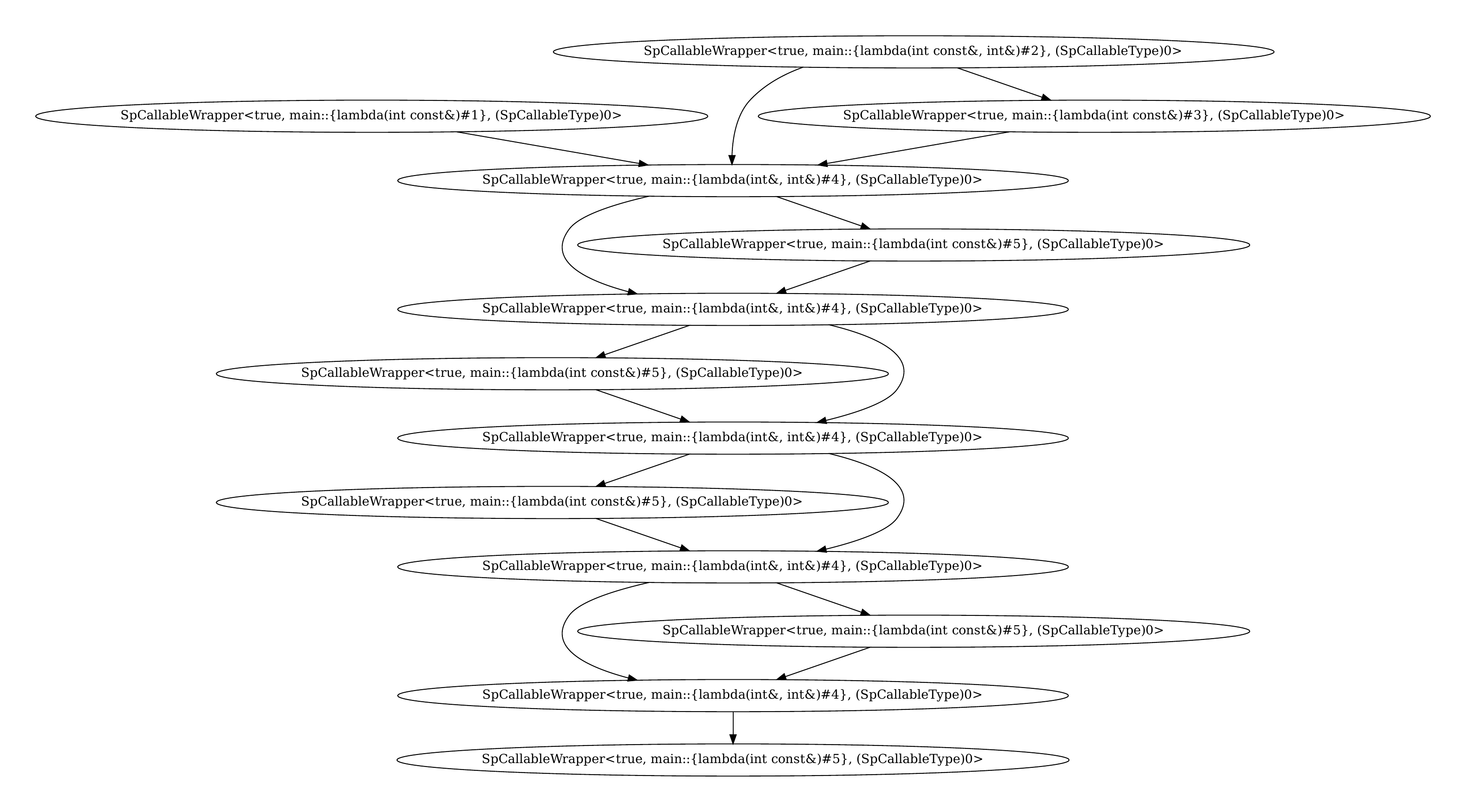}
        \caption{Task graph of the \emph{daggraph} example, generated from the dot file.}
        \label{fig:vizu:graph}
    \end{subfigure}
    \hfill
    \begin{subfigure}{0.47\textwidth}
        \centering
        \includegraphics[width=\textwidth]{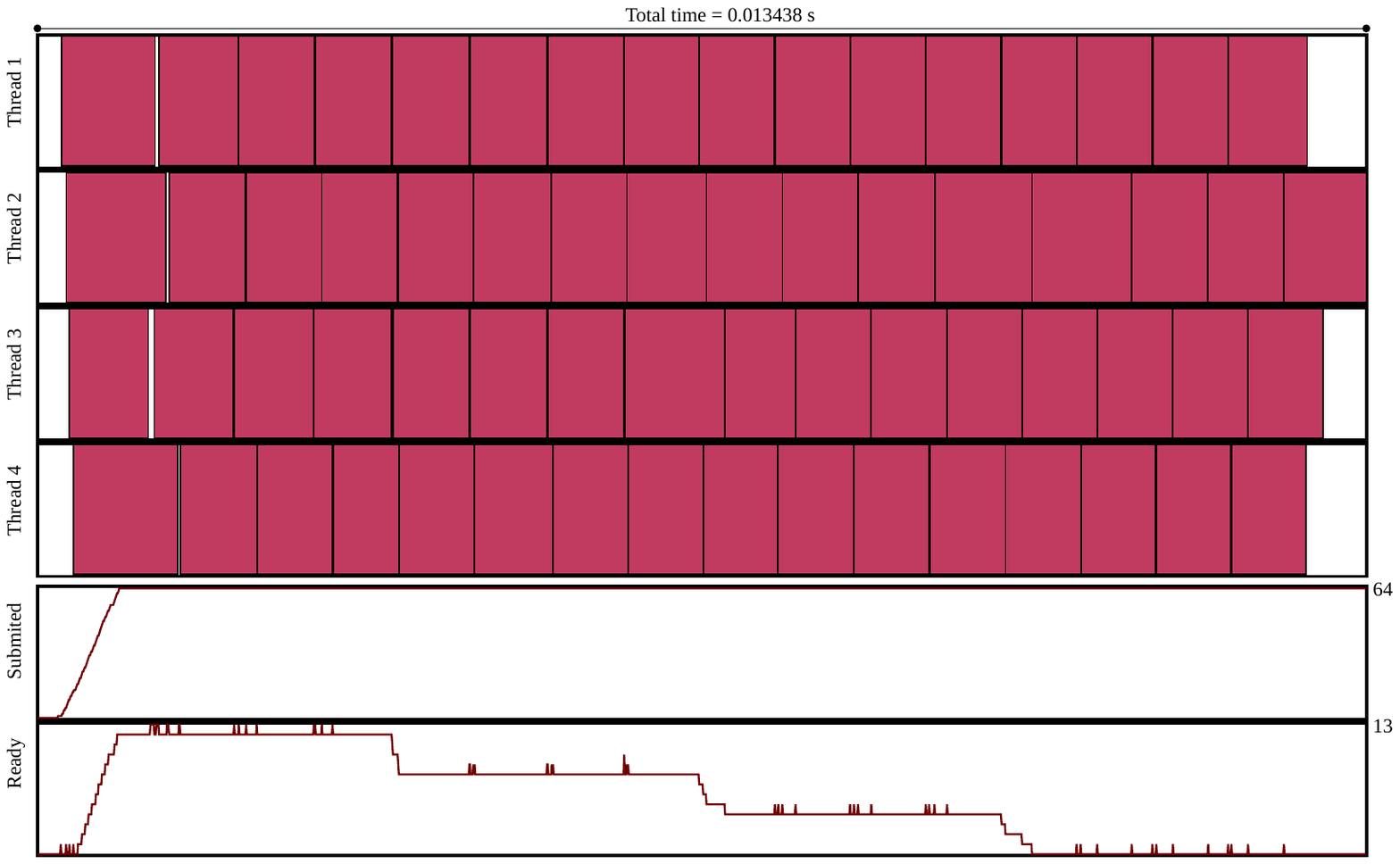}
        \caption{Execution trace of the GEMM test case using 4 threads, a matrix of size 512 and blocks of size 128.}
        \label{fig:vizu:trace}
    \end{subfigure}
    \caption{Example of graphs and execution trace exported after a run.}
    \label{fig:vizu}
\end{figure}

A graph and an execution trace are provided in Figure~\ref{fig:vizu}, and the corresponding calls are given in Code~\ref{code:vizu}.

\begin{figure}[h!]
\begin{lstlisting}[escapechar=|, label={code:vizu}, caption={Specx example - export of the task graph and the execution trace.}]
// Export the graph
tg.generateDot("/tmp/graph.dot");
// Export the execution trace (false means: hide the dependencies)
tg.generateTrace("/tmp/gemm-simu.svg", false);
\end{lstlisting}
\end{figure}

\section{Performance and usability study}
\label{sec:study}

\subsection{Overhead estimation}

In this section, we aim to estimate the overhead of using the Specx runtime system.

\paragraph{Configuration}
We assess our method on the following configuration:
\begin{itemize}
    \item Intel-AVX512: it is a 2 $\times$ 18-core Cascade Lake Intel Xeon Gold 6240 at 2.6 GHz with AVX-512 (Advanced Vector 512-bit,  Foundation,  Conflict Detection, Byte and Word, Doubleword and Quadword Instructions, and Vector Length).
    The main memory consists of 190 GB DRAM memory arranged in two NUMA nodes. 
    Each CPU has 18 cores with 32KB private L1 cache, 1024KB private L2 cache, and 25MB shared L3 cache.
    We use the GNU compiler 11.2.0 and the MKL 2022.0.2.
\end{itemize}

\paragraph{Results}

In this section, we discuss the engine overhead that we evaluate with the following pattern.
We create a runtime system with $T$ CPU workers and $T$ distinct data objects.
Then, we insert $T \times N$ tasks, with each task accessing one of the data object.
Consequently, the task graph we generate is actually composed of $T$ independent sub-graphs.
Inside each task, the worker that execute it, simply waits for a given duration $D$.
As a result, the final execution time is given $N \times (D + O)$, where $O$ is an overhead of picking a task from the runtime.
Also, we can measure the time it takes to insert the $T \times N$ tasks to obtain an insertion cost $I$.

\begin{figure}[h]
    \centering    

    \begin{subfigure}{\textwidth}
        \centering
        \includegraphics[width=.7\textwidth,trim={6 8cm 0 0},clip]{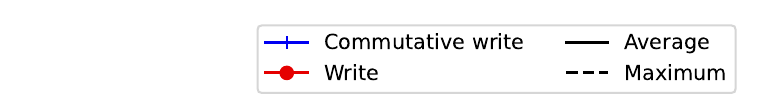}
    \end{subfigure}
    
    \begin{subfigure}{0.47\textwidth}
        \centering
        \includegraphics[width=\textwidth]{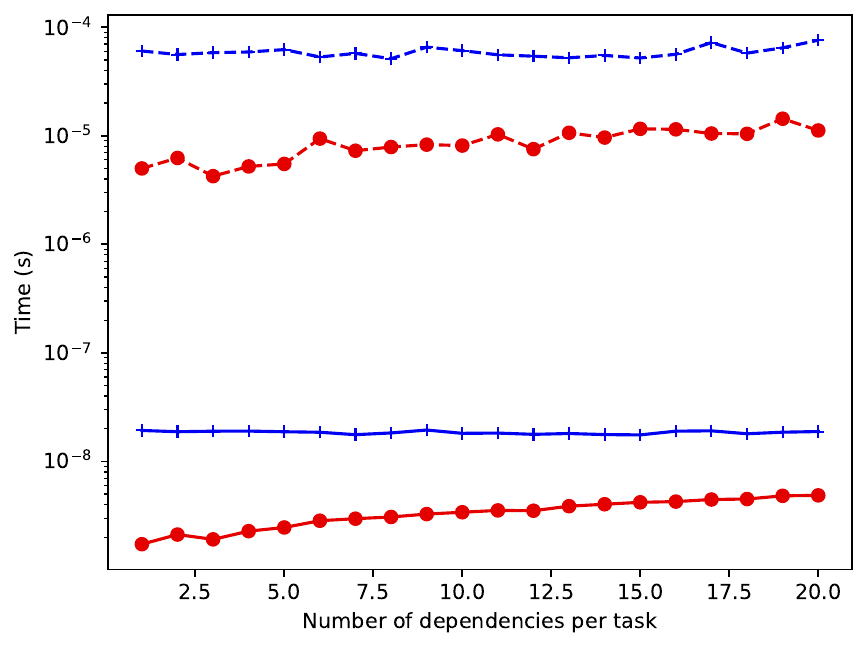}
        \caption{Insertion - $D = 10^{-5}$}
        \label{fig:overhead:insertion-5}
    \end{subfigure}
    \hfill
    \begin{subfigure}{0.47\textwidth}
        \centering
        \includegraphics[width=\textwidth]{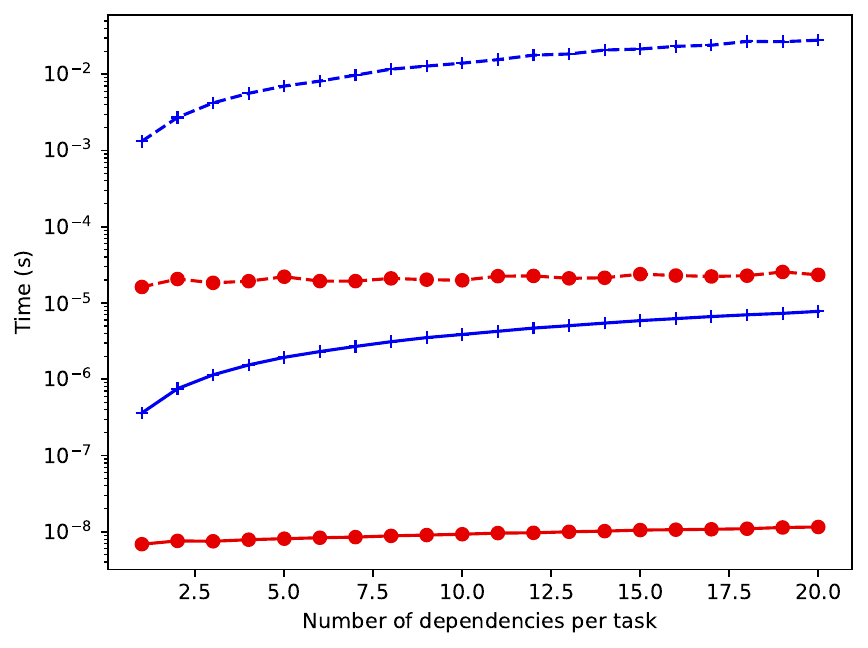}
        \caption{Per task - $D = 10^{-5}$}
        \label{fig:overhead:overhead-5}
    \end{subfigure}
    
    \begin{subfigure}{0.47\textwidth}
        \centering
        \includegraphics[width=\textwidth]{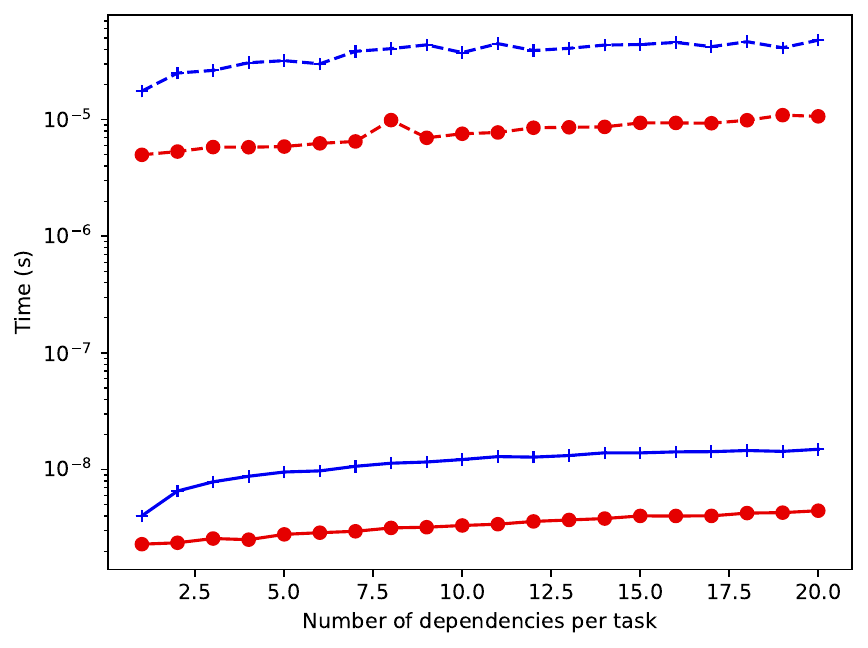}
        \caption{Insertion - $D = 10^{-4}$}
        \label{fig:overhead:insertion-4}
    \end{subfigure}
    \hfill
    \begin{subfigure}{0.47\textwidth}
        \centering
        \includegraphics[width=\textwidth]{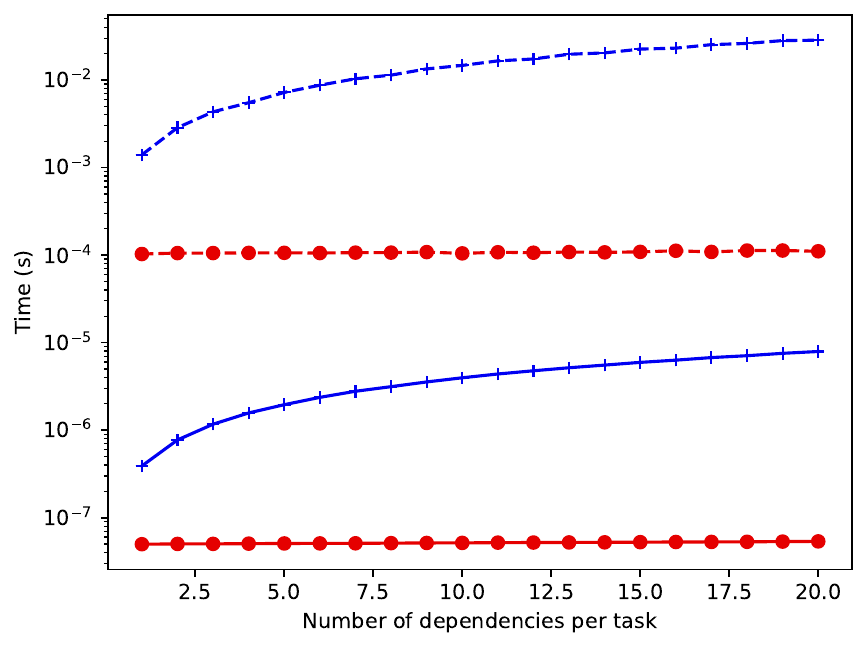}
        \caption{Per task - $D = 10^{-4}$}
        \label{fig:overhead:overhead-4}
    \end{subfigure}
    
    \begin{subfigure}{0.47\textwidth}
        \centering
        \includegraphics[width=\textwidth]{overhead/gcc_insertion_0_001}
        \caption{Insertion - $D = 10^{-3}$}
        \label{fig:overhead:insertion-3}
    \end{subfigure}
    \hfill
    \begin{subfigure}{0.47\textwidth}
        \centering
        \includegraphics[width=\textwidth]{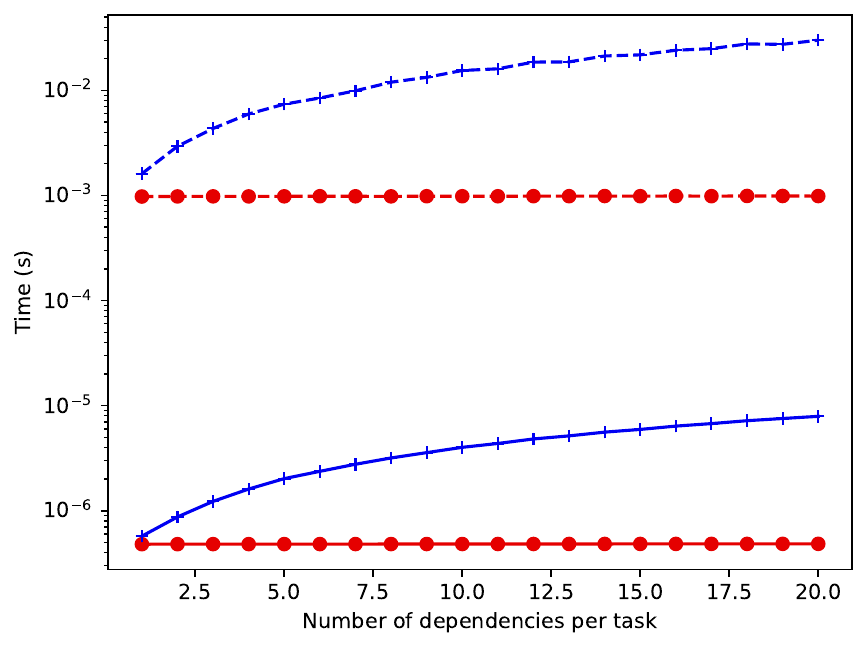}
        \caption{Per task - $D = 10^{-3}$}
        \label{fig:overhead:clang-3}
    \end{subfigure}
    
    \caption{Estimation of the overheads for the \emph{write} ($\bullet$) and \emph{commutative-write} ($+$) data accesses for different number of dependencies.
    We provide the maximum overhead reached ($- -$) and the average one ($-$).
             The overhead is given for picking a task $O$ (right column) and the insertion $I$ (left column).}
    \label{fig:overhead}
\end{figure}

We provide the result in Figure~\ref{fig:overhead} for 1 to 20 dependencies.
As expected, the overhead of using commutative write is significant compared to a normal write.
The insertion cost is also higher when the tasks duration is smaller ($D=10^{-5}$).
The reason is that as the tasks are smaller, the worker query to the runtime system more often, which can compete with the insertion of ready tasks by the master thread and create contention.
The cost also increase slightly as the number of dependencies in the tasks increases.
Finally, the overhead per tasks is stable as the number of dependencies per tasks increases for the write access.
However, for the commutative access, the overhead increases with the number of dependencies.

\subsection{Numerical Applications}

In this section, we evaluate the performance of Specx on four numerical applications.

\paragraph{Test cases:}
\begin{itemize}
    \item Vector-scalar product (Blas axpy): 
    In this application, we compute the axpy operation of a vector that is split into pieces. 
    We vary the size of these pieces (the number of values computed in each task) and the number of pieces (the number of tasks). 
    All tasks have the same computational cost and are independent. 
    Each memory block is used only once, either for reading or writing. 
    The kernels are implemented using our own code.
    We use simple floating-point precision.
    \item General matrix-matrix product (Blas Gemm):
    In this application, we split three square matrices of the same dimension into blocks/panels of the same size. 
    We vary both the dimension of the matrices and the size of the blocks. 
    Consequently, all tasks have the same computational cost, though there are dependencies on the blocks of the output matrix (usually refered as $C$). 
    The tasks call Blas functions (Intel MKL on CPU and cuBlas on GPU).
    We use double floating-point precision.
    \item Cholesky factorization:
    In this application, we split a matrix into blocks/panels and perform a Cholesky factorization by calling Blas/Lapack functions, 
    which results in four types of tasks (\texttt{POTRF}, \texttt{TRSM}, \texttt{SYRK}, and \texttt{GEMM}). 
    All tasks of the same type have the same computational cost. 
    Each task has predecessors and/or successors, and some tasks are higly critical, i.e., they release lots of tasks or are on the critical path. 
    The tasks call Blas/Lapack functions (Intel MKL on CPU and cuBlas on GPU).
    We use double floating-point precision.
    \item Particle interactions (n-body):
    In this application, we generate groups of particles and perform two types of computations: 
    inner (interactions of particles within the same block) and outer (interactions between two different groups of particles). 
    We vary the number of blocks and the number of particles in each group. 
    As a result, most of the tasks have different computational costs. 
    In our implementation, dependencies are applied to complete groups; we do not split the groups into symbolic data and computational data.
    The kernel is the computation of a direct summation or coulomb potential, giving around 17 Flops and one SQRT per interaction.
    We use double floating-point precision.
\end{itemize}

We remind that in our study, we are not interested in the performance of the kernels, but in the performance of the runtime system.
We want to evaluate the overhead of the runtime system, the performance of the task-based approach, the scheduling choices, and the data management.

\paragraph{Configurations:}

We use the following software configuration:
NVCC 12.3, the GNU compiler 10.2.0 and Intel MKL 2020.

We tested on four different hardware platforms:
\begin{itemize}
    \item \textbf{rtx8000} : 2 NVIDIA Quadro RTX8000 GPUs (48GB) and 2x 20-core Cascade Lake Intel Xeon Gold 5218R CPU at 2.10GHz (190GB).
                            SM 7.5 (Turing architecture)
    \item \textbf{v100} : 2 NVIDIA V100 GPUs (16GB) and 2x 16-core Skylake Intel Xeon Gold 6142 at 2.6GHz (380GB).
                            SM 7.0 (Volta architecture)
    \item \textbf{p100} : 2 NVIDIA P100 GPUs (16GB) and 2x 16-core Broadwell Intel Xeon E5-2683 v4 at 2.1GHz (128GB).
                            SM 6.0 (Pascal architecture)
    \item \textbf{a100} : 2 NVIDIA A100 GPUs (40GB) and 2x 24-core AMD Zen2 EPYC 7402 at 2.80GHz (512GB).
                            SM 8.0 (Ampere architecture)
\end{itemize}

For each GPU we use four streams, which means that we create four CPU threads with each thread having a stream and the four threads sharing the same GPU.

\paragraph{Results}

\begin{figure}[h]
    \centering    

    \begin{subfigure}{\textwidth}
        \centering
        \includegraphics[width=.5\textwidth]{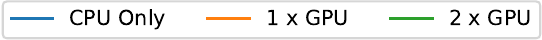}
    \end{subfigure}
    
    \begin{subfigure}{0.47\textwidth}
        \centering
        \includegraphics[width=\textwidth]{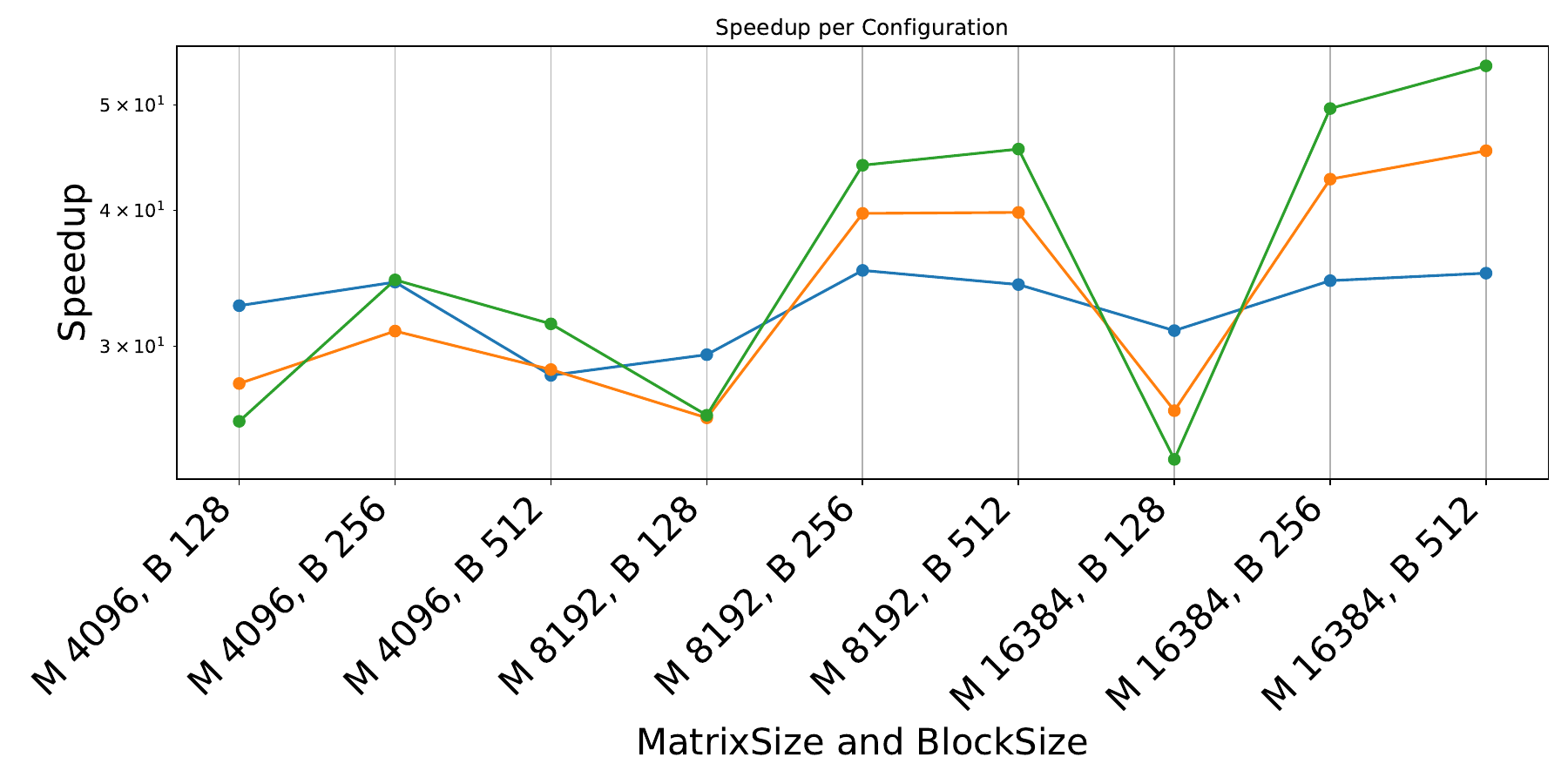}
        \caption{RTX GEMM}
        \label{fig:rtx:gemm}
    \end{subfigure}
    \hfill
    \begin{subfigure}{0.47\textwidth}
        \centering
        \includegraphics[width=\textwidth]{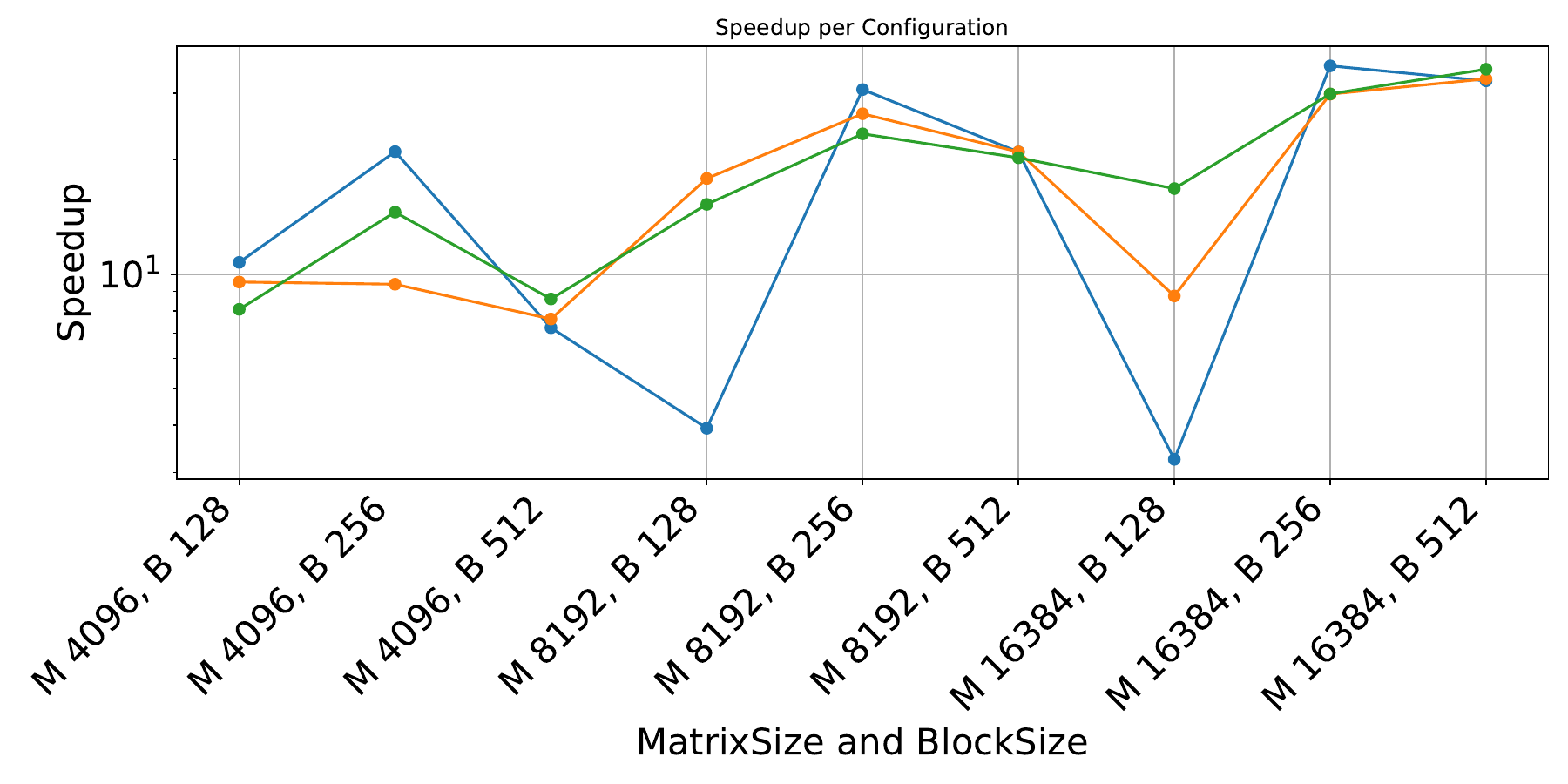}
        \caption{RTX Cholesky}
        \label{fig:rtx:cholesky}
    \end{subfigure}
    
    \begin{subfigure}{0.47\textwidth}
        \centering
        \includegraphics[width=\textwidth]{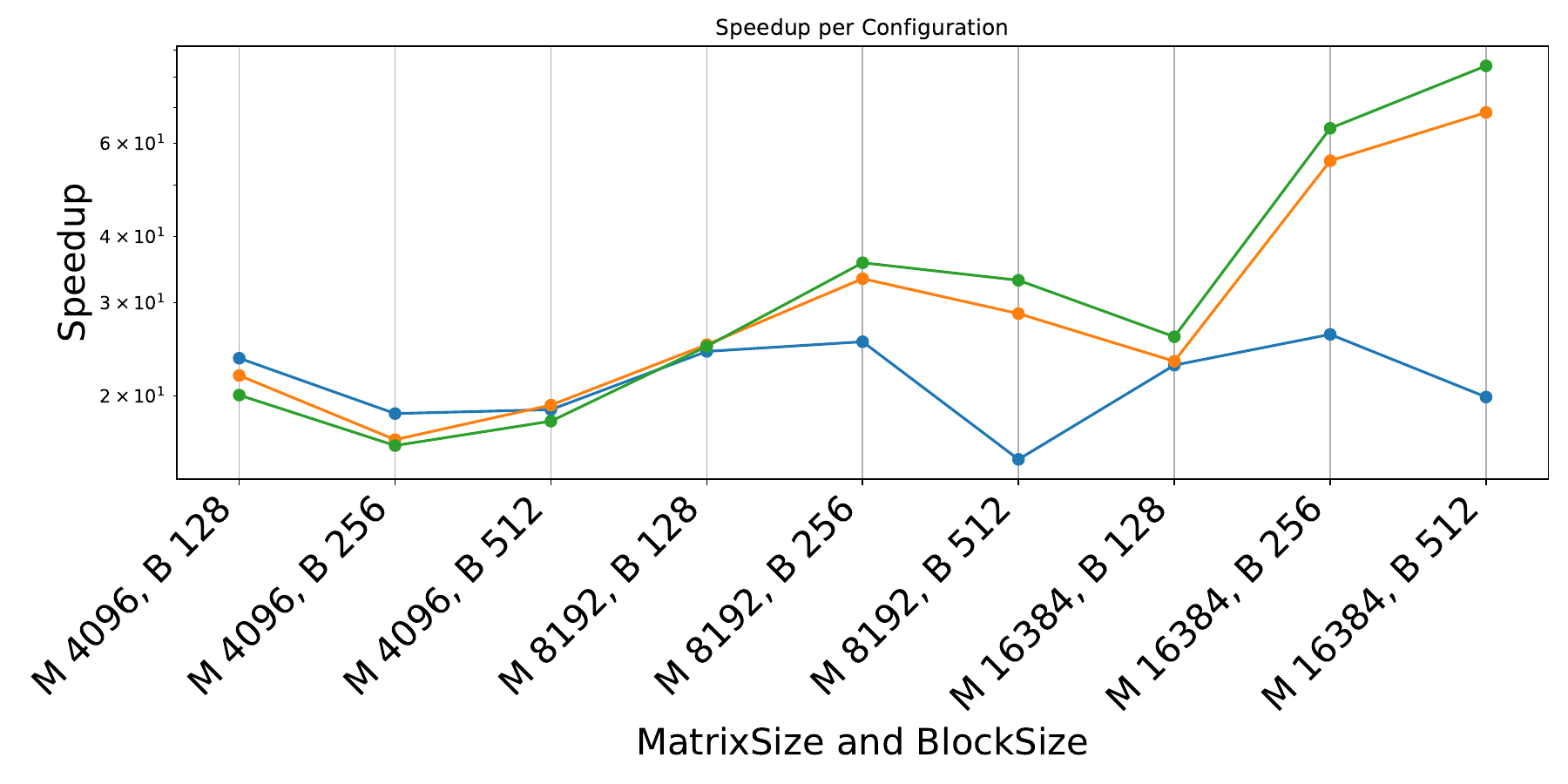}
        \caption{P100 GEMM}
        \label{fig:P100:gemm}
    \end{subfigure}
    \hfill
    \begin{subfigure}{0.47\textwidth}
        \centering
        \includegraphics[width=\textwidth]{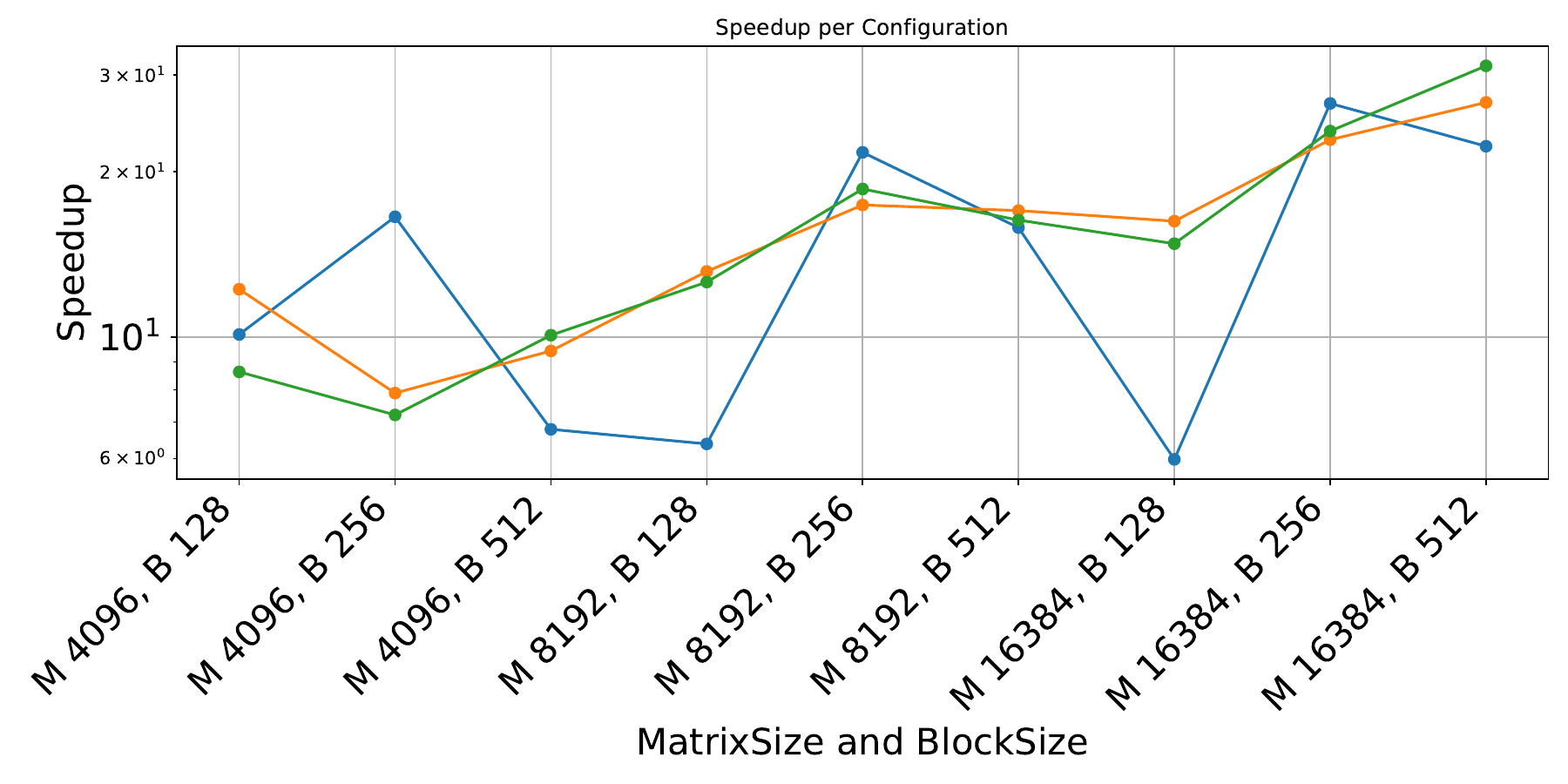}
        \caption{P100 Cholesky}
        \label{fig:P100:cholesky}
    \end{subfigure}
    
    \begin{subfigure}{0.47\textwidth}
        \centering
        \includegraphics[width=\textwidth]{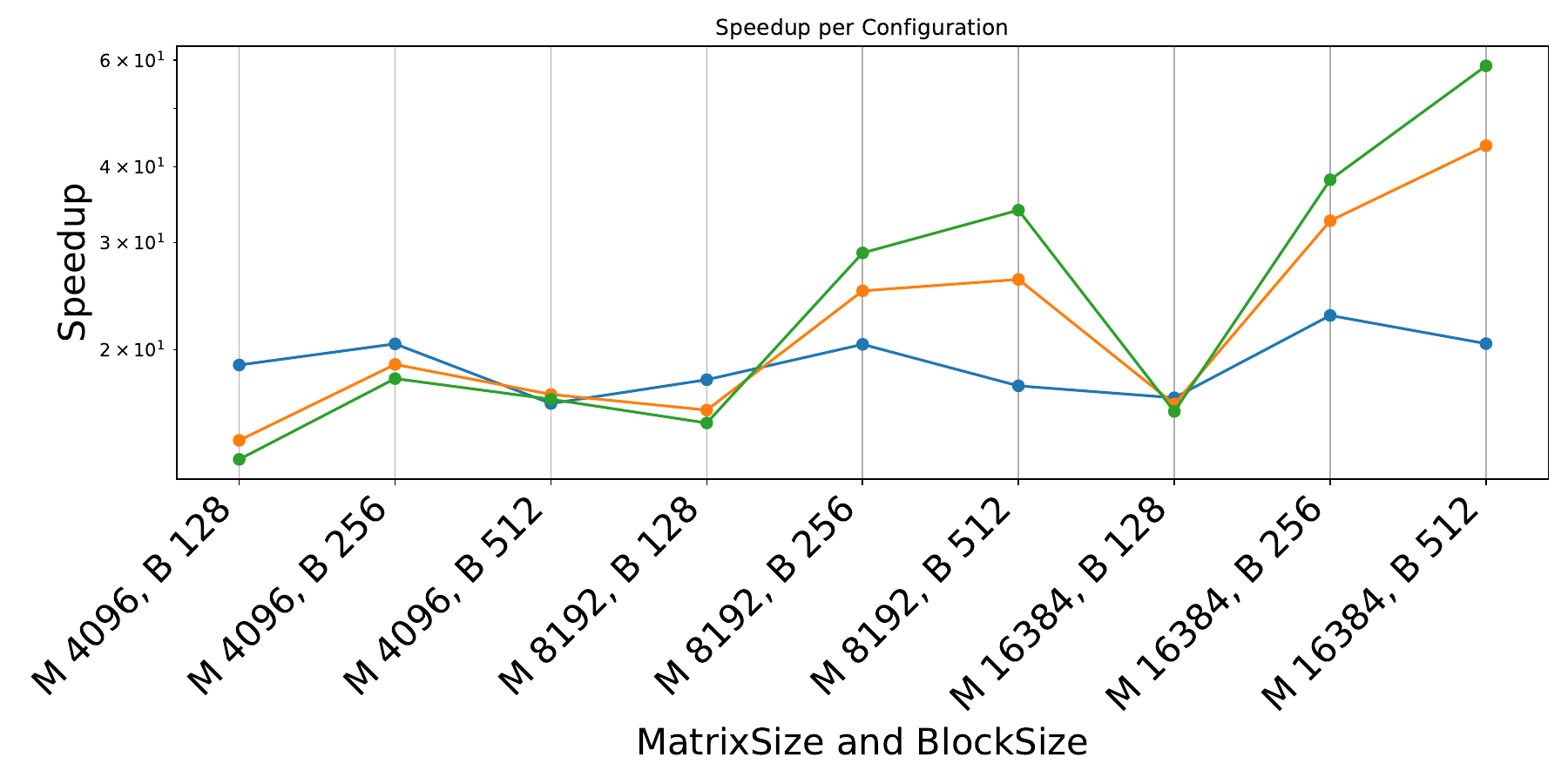}
        \caption{V100 GEMM}
        \label{fig:V100:gemm}
    \end{subfigure}
    \hfill
    \begin{subfigure}{0.47\textwidth}
        \centering
        \includegraphics[width=\textwidth]{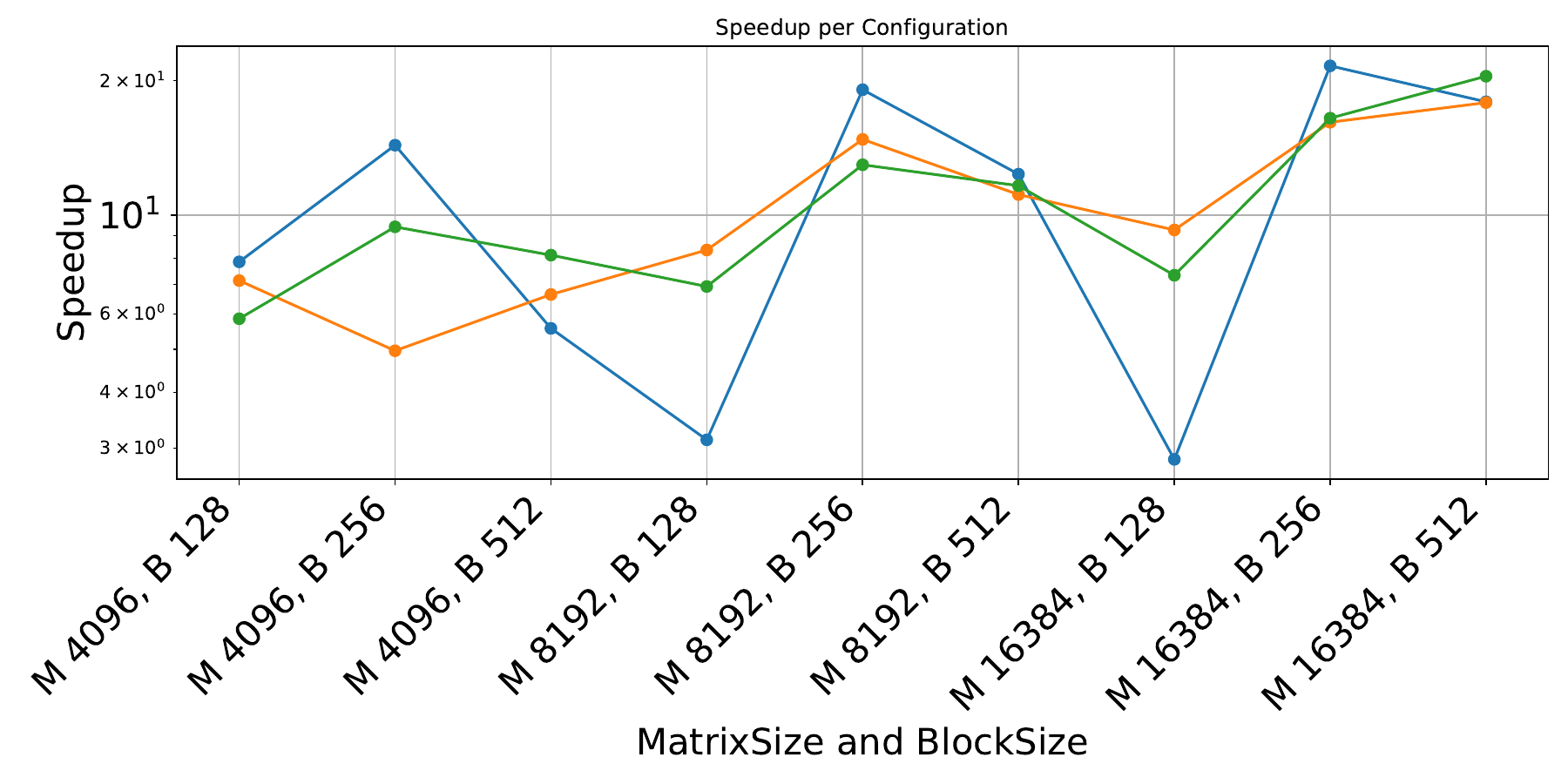}
        \caption{V100 Cholesky}
        \label{fig:V100:cholesky}
    \end{subfigure}
    
    \begin{subfigure}{0.47\textwidth}
        \centering
        \includegraphics[width=\textwidth]{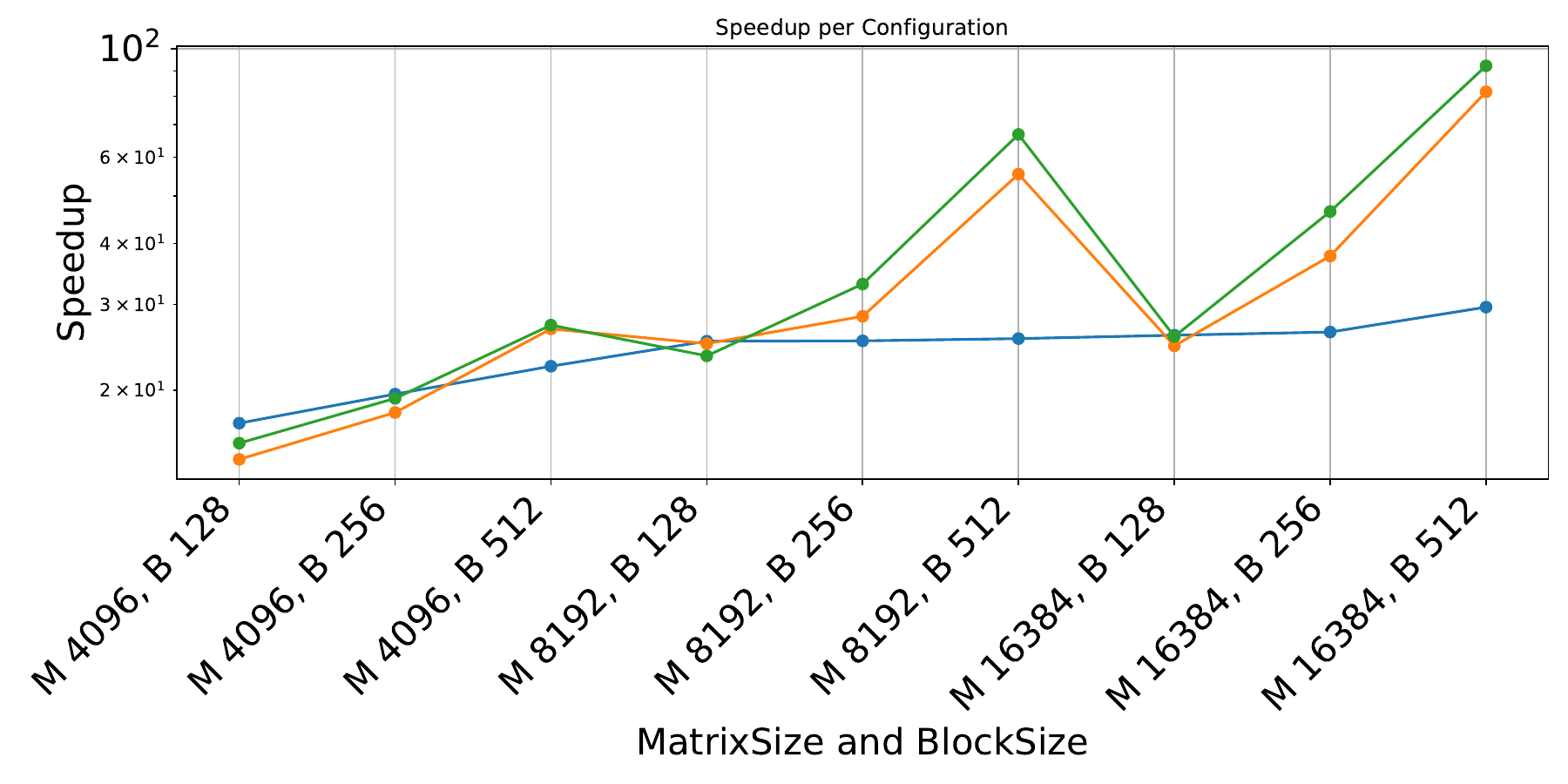}
        \caption{A100 GEMM}
        \label{fig:A100:gemm}
    \end{subfigure}
    \hfill
    \begin{subfigure}{0.47\textwidth}
        \centering
        \includegraphics[width=\textwidth]{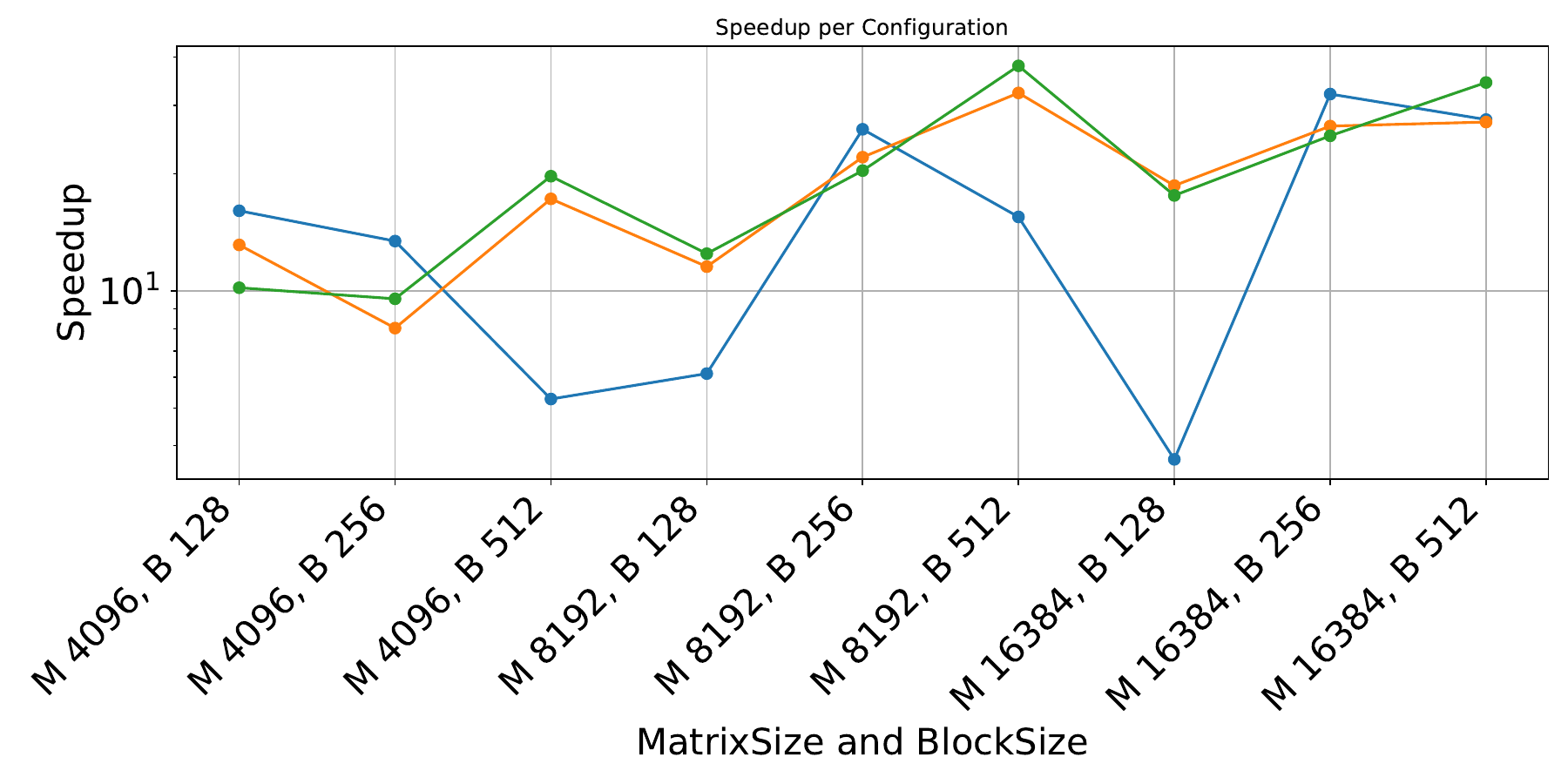}
        \caption{A100 Cholesky}
        \label{fig:A100:cholesky}
    \end{subfigure}
    
    \caption{Performance results for the GEMM and Cholesky test cases.
             The $x$-axis represents the test case's size, and the $y$-axis represents the speedup over a sequential execution.}
    \label{fig:gemm_cholesky}
\end{figure}

\begin{figure}[h]
    \centering    

    \begin{subfigure}{\textwidth}
        \centering
        \includegraphics[width=.5\textwidth]{Results/gpu_legend}
    \end{subfigure}
    
    \begin{subfigure}{0.47\textwidth}
        \centering
        \includegraphics[width=\textwidth]{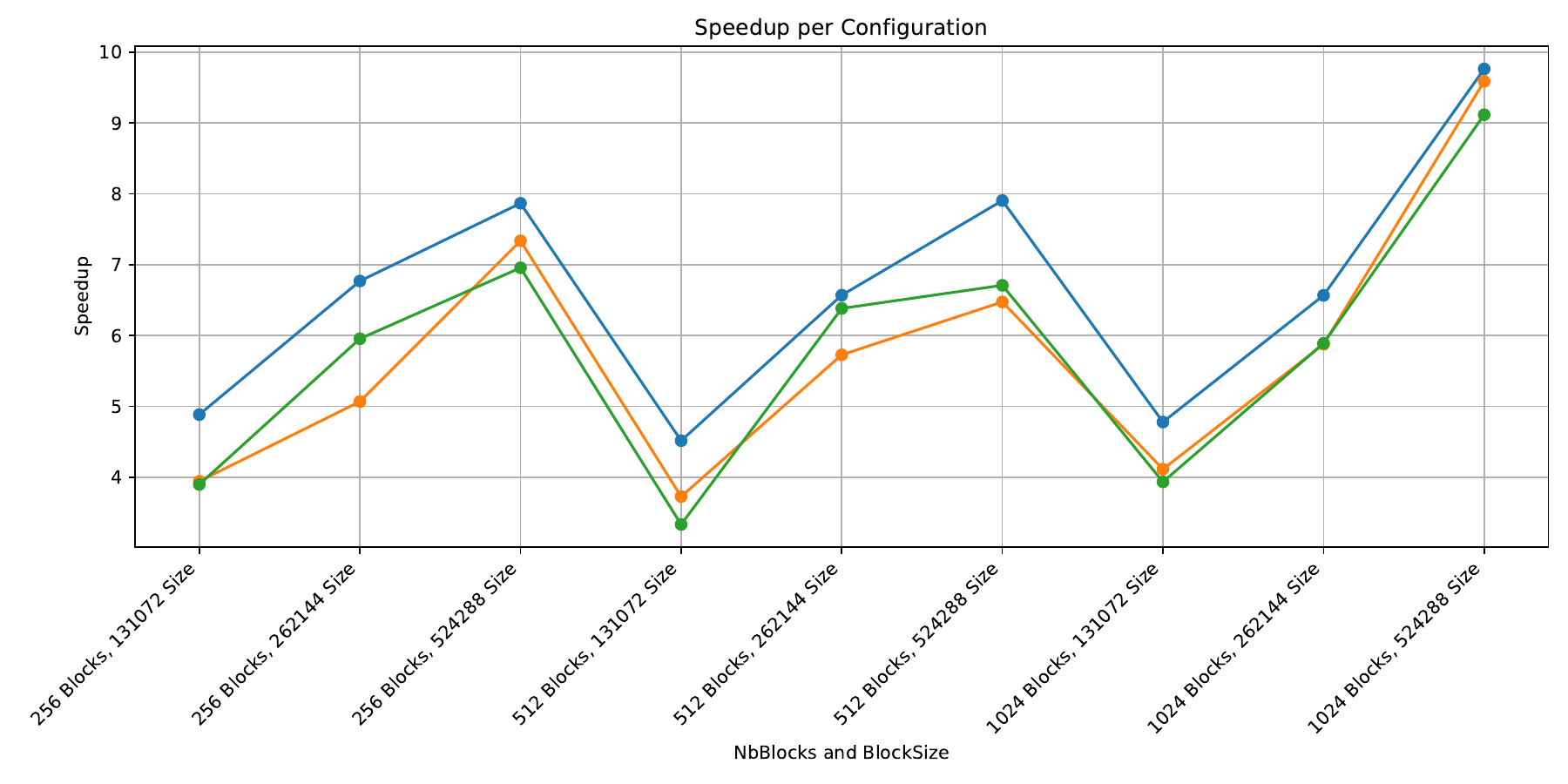}
        \caption{RTX AXPY}
        \label{fig:rtx:axpy}
    \end{subfigure}
    \hfill
    \begin{subfigure}{0.47\textwidth}
        \centering
        \includegraphics[width=\textwidth]{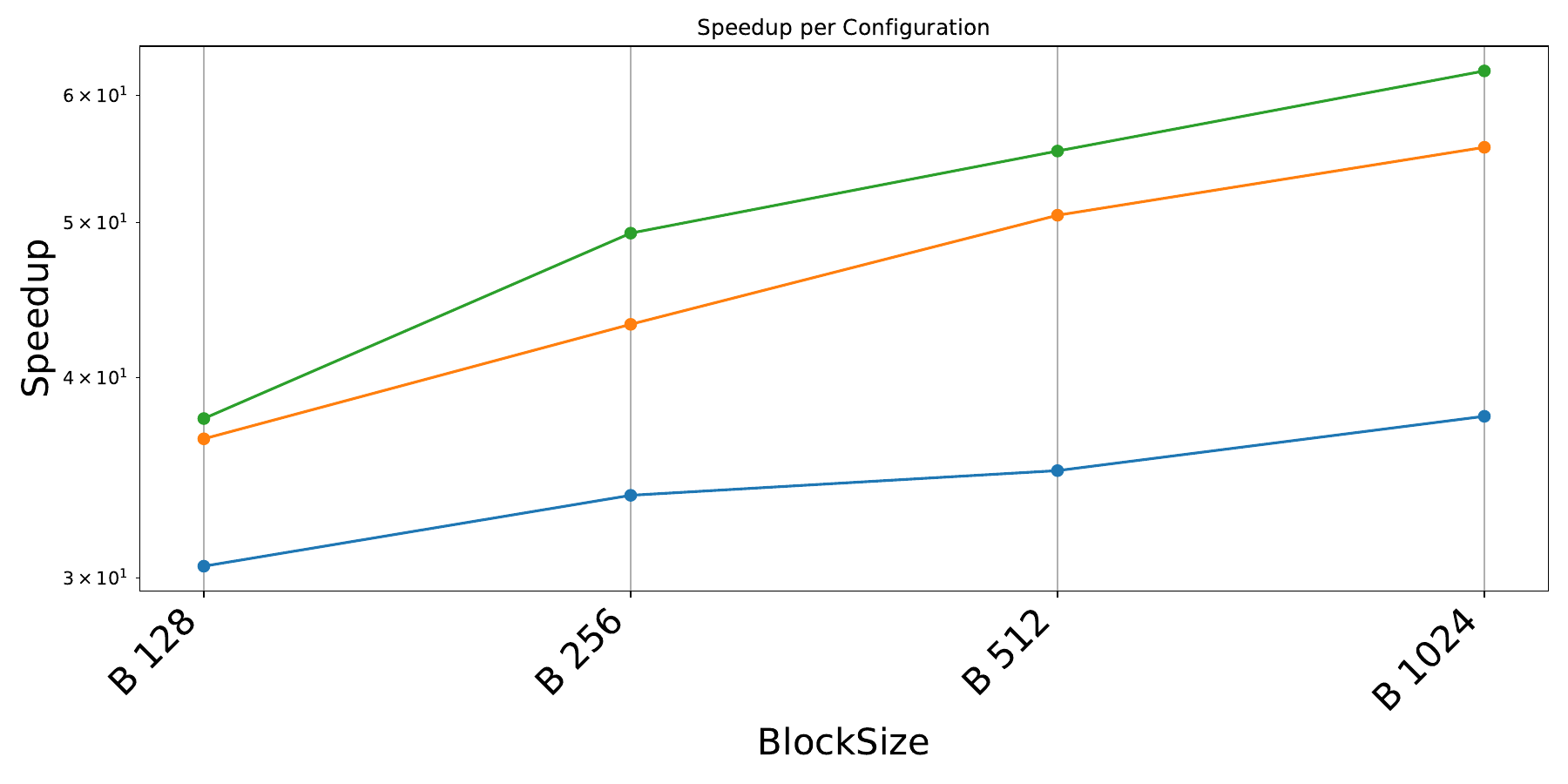}
        \caption{RTX Particle simulation}
        \label{fig:rtx:particles}
    \end{subfigure}
    
    \begin{subfigure}{0.47\textwidth}
        \centering
        \includegraphics[width=\textwidth]{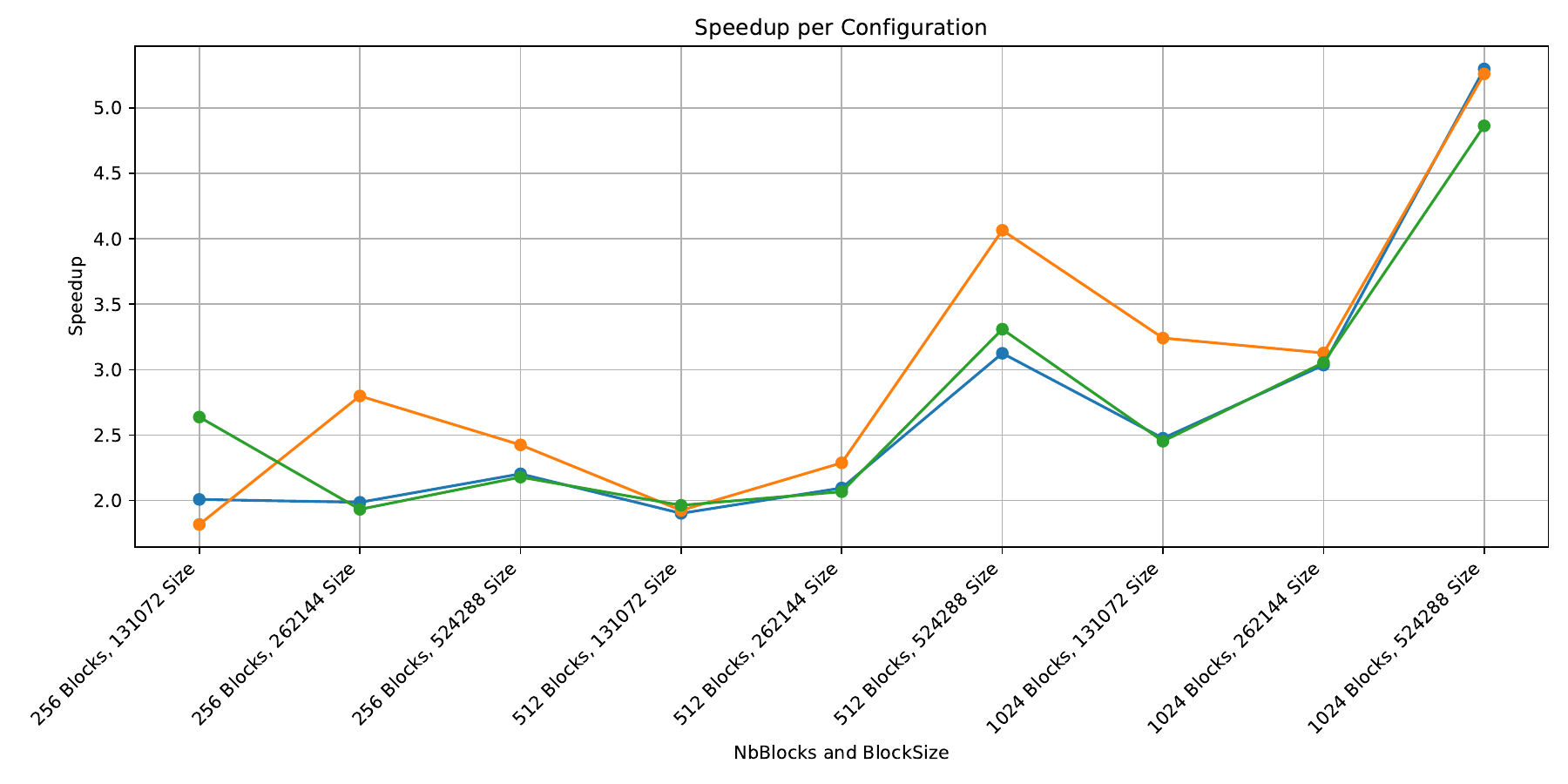}
        \caption{P100 AXPY}
        \label{fig:P100:axpy}
    \end{subfigure}
    \hfill
    \begin{subfigure}{0.47\textwidth}
        \centering
        \includegraphics[width=\textwidth]{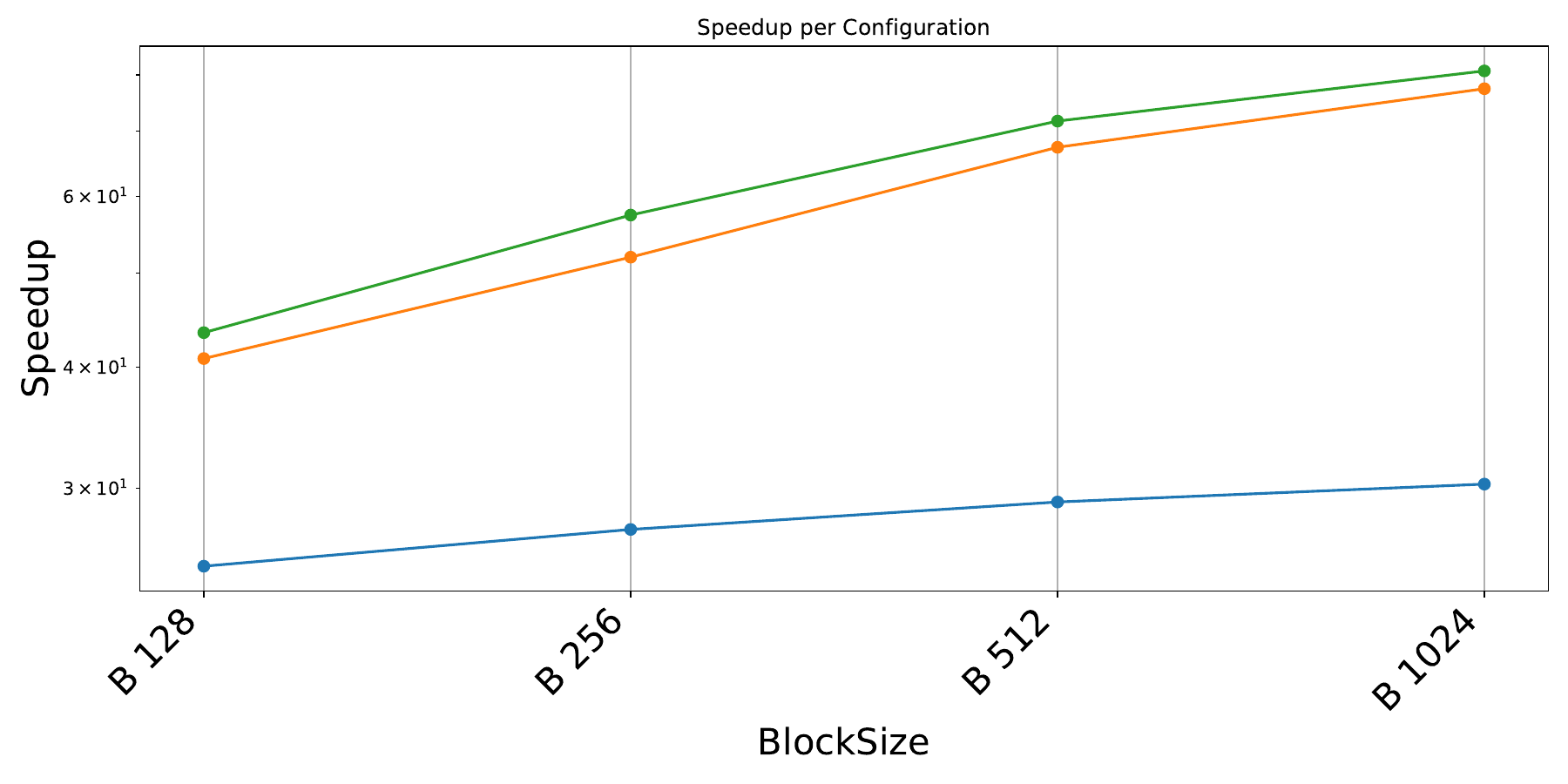}
        \caption{P100 Particle simulation}
        \label{fig:P100:particles}
    \end{subfigure}
    
    \begin{subfigure}{0.47\textwidth}
        \centering
        \includegraphics[width=\textwidth]{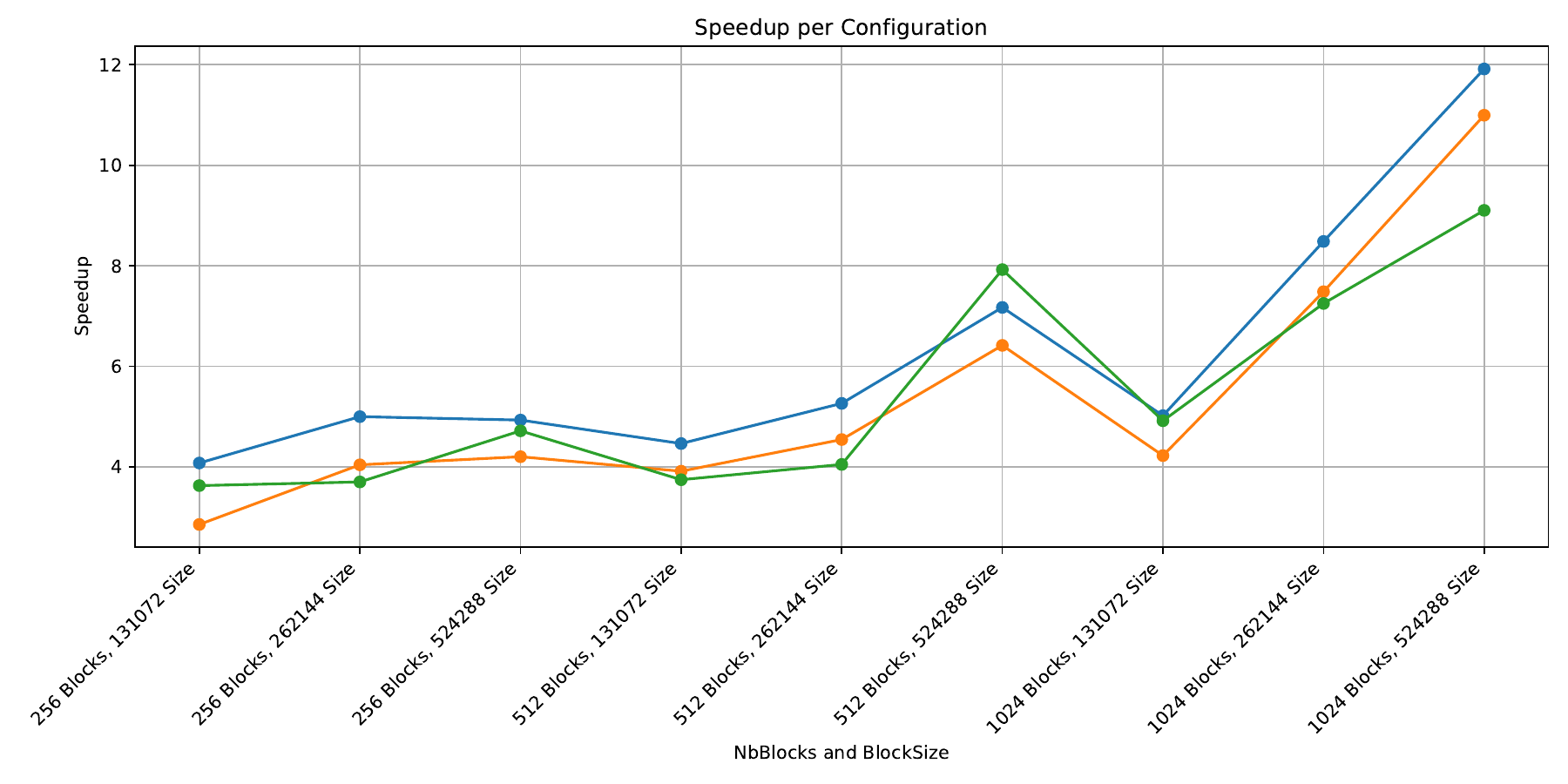}
        \caption{V100 AXPY}
        \label{fig:V100:axpy}
    \end{subfigure}
    \hfill
    \begin{subfigure}{0.47\textwidth}
        \centering
        \includegraphics[width=\textwidth]{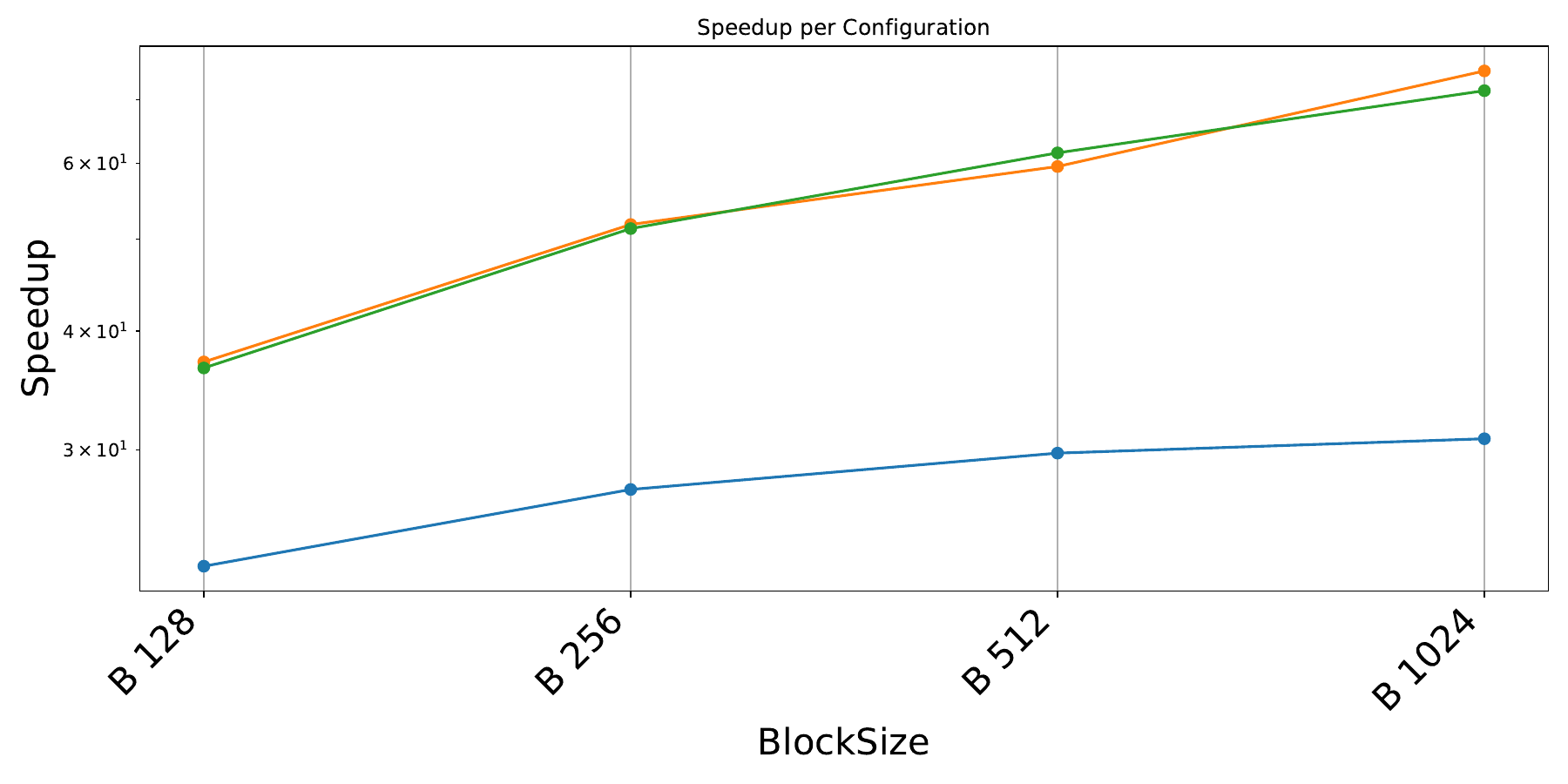}
        \caption{V100 Particle simulation}
        \label{fig:V100:particles}
    \end{subfigure}
    
    \begin{subfigure}{0.47\textwidth}
        \centering
        \includegraphics[width=\textwidth]{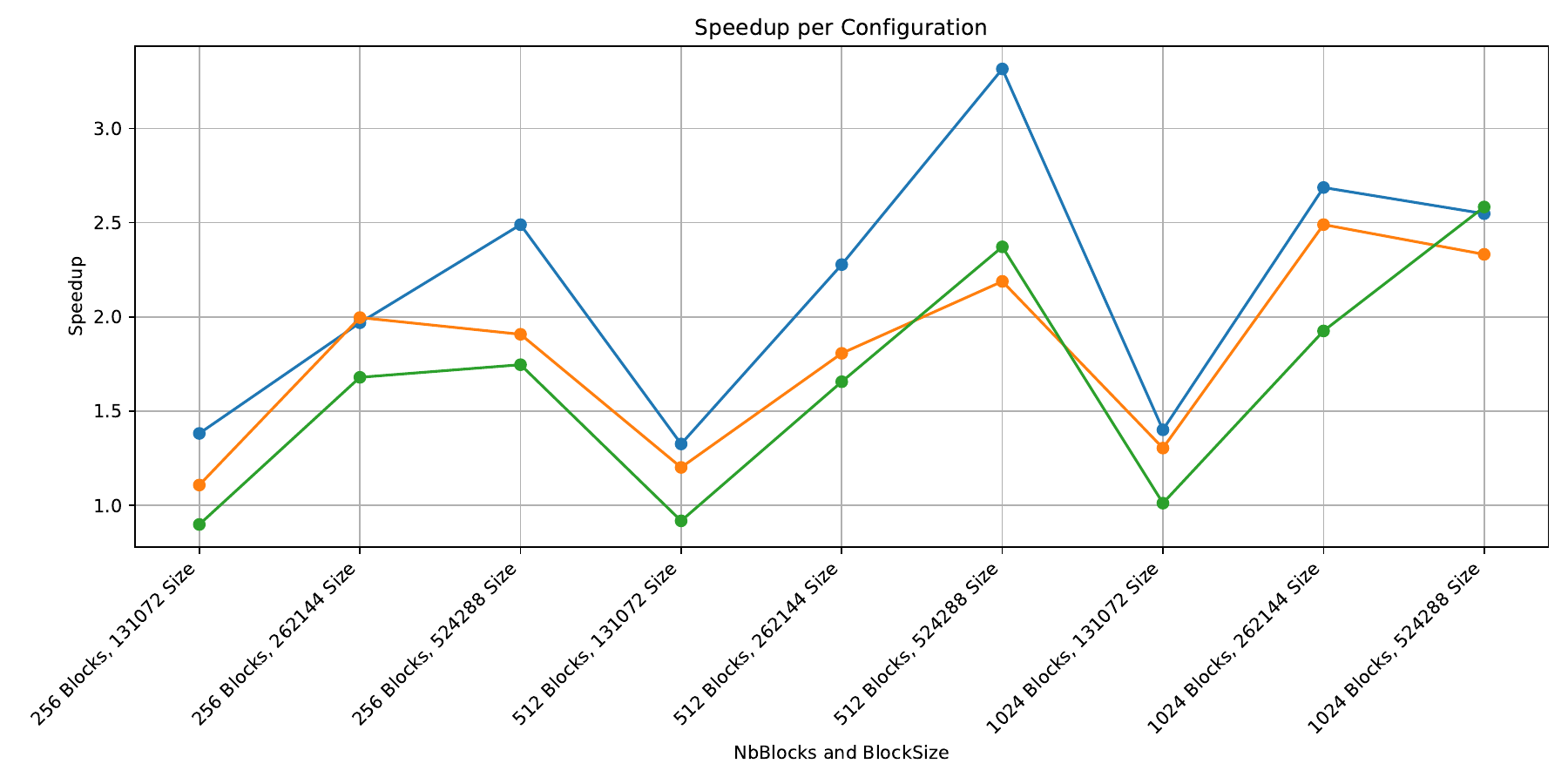}
        \caption{A100 AXPY}
        \label{fig:A100:axpy}
    \end{subfigure}
    \hfill
    \begin{subfigure}{0.47\textwidth}
        \centering
        \includegraphics[width=\textwidth]{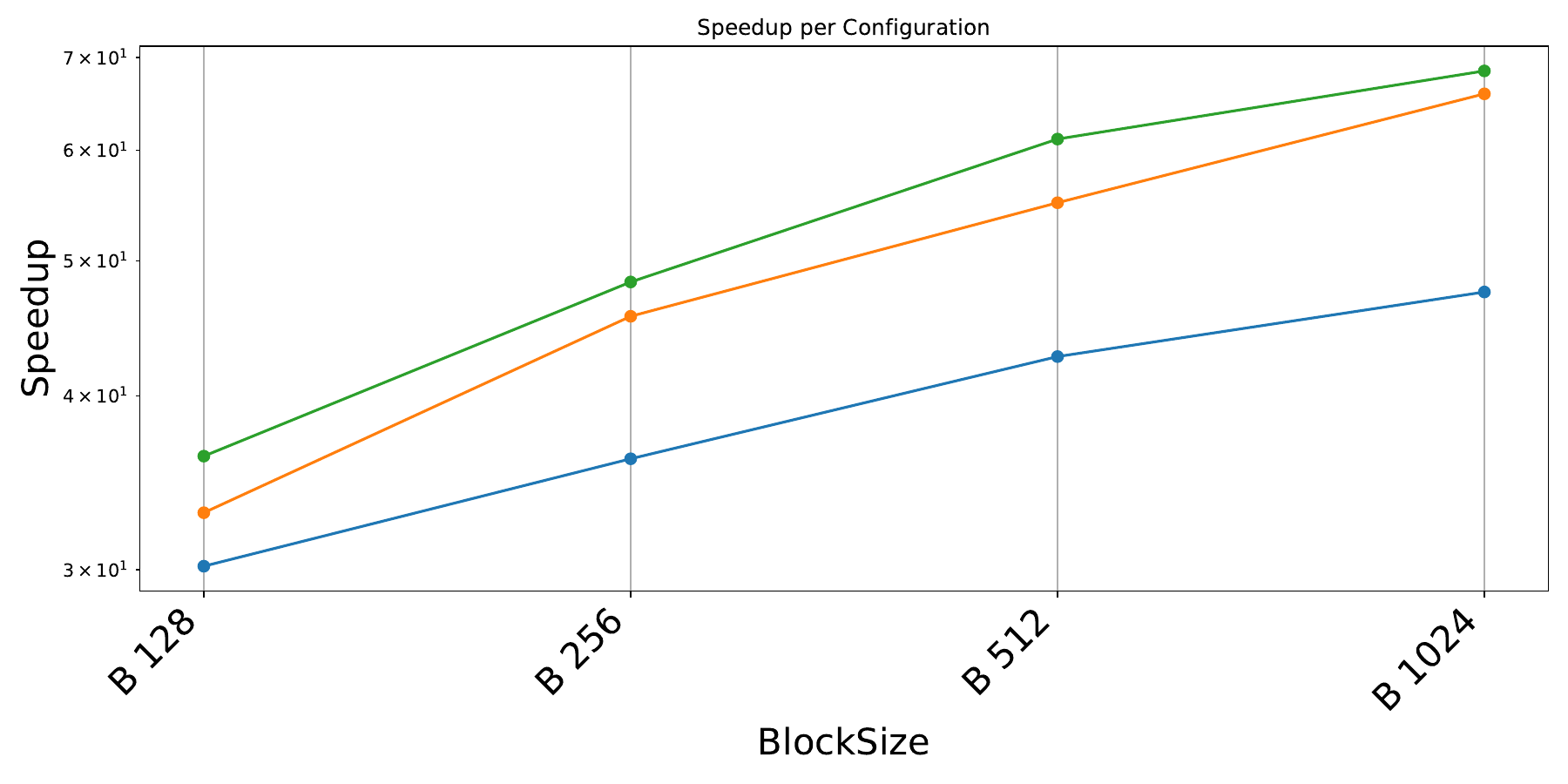}
        \caption{A100 Particle simulation}
        \label{fig:A100:particles}
    \end{subfigure}
    
    \caption{Performance results for the axpy and particle test cases.
             The $x$-axis represents the test case's size, and the $y$-axis represents the speedup over a sequential execution.}
    \label{fig:axpy_particles}
\end{figure}

In Figure~\ref{fig:gemm_cholesky}, we present the results for the GEMM and Cholesky test cases, 
while Figure~\ref{fig:axpy_particles} provides the results for the AXPY and particle interaction test cases.

The results indicate that the performance of the GEMM test cases is significantly influenced by the sizes of the blocks and matrices. 
GPUs show a substantial speedup for matrices with dimensions of 16,384, particularly with block sizes larger than 128. 
This demonstrates that when GPUs do not have enough work to process within a task, the overhead associated with task management can become significant, sometimes leading to performance lower than that of the CPU. 
This effect is even more pronounced in the AXPY test case, where the CPU outperforms the GPU for most configurations.

Another notable effect is the poor execution performance near the end of certain configurations. 
For instance, in the GEMM test on the P100 GPU (Figure~\ref{fig:P100:gemm}), the CPU experiences a performance drop for the 8192/512 configuration. 
In this scenario, the workload per task is substantial, and when only one or a few threads remain active to compute the final tasks, the rest of the threads remain idle, leading to a dramatic decline in speedup.

The Cholesky test cases exhibit highly variable performance.
When using GPUs, the performance appears more stable, though the speedup is not as significant as in the GEMM test case. 
For CPU-only configurations, performance is highly inconsistent, with notable drops for block sizes of 128, among others. 
These variations are attributable to scheduling decisions, which does not take into account the criticality in its decisions.
Consequently, a good execution order is based on luck, and the performance can be highly variable.

In the particle simulations, GPUs provide a significant speedup.
However, the benefit of using two GPUs instead of one is negligible. 
For example, on the RTX GPUs (Figure~\ref{fig:rtx:particles}), the speedup is approximately 53 for one GPU and 60 for two GPUs when compared to sequential execution.

\newcommand{\archrtx}{./Results/build-75/results--20241103_151509/}
\newcommand{\archv}{./Results/build-70/results--20241103_151359/}
\newcommand{\archp}{./Results/build-60/results--20241103_163144/}
\newcommand{\archa}{./Results/build-80/results--20241103_222447/}

\begin{figure}[h]
    \centering    

    \begin{subfigure}{\textwidth}
        \centering
        \includegraphics[width=.5\textwidth]{Results/gpu_legend}
    \end{subfigure}
    
    \begin{subfigure}{0.30\textwidth}
        \centering
        \includegraphics[width=\textwidth]{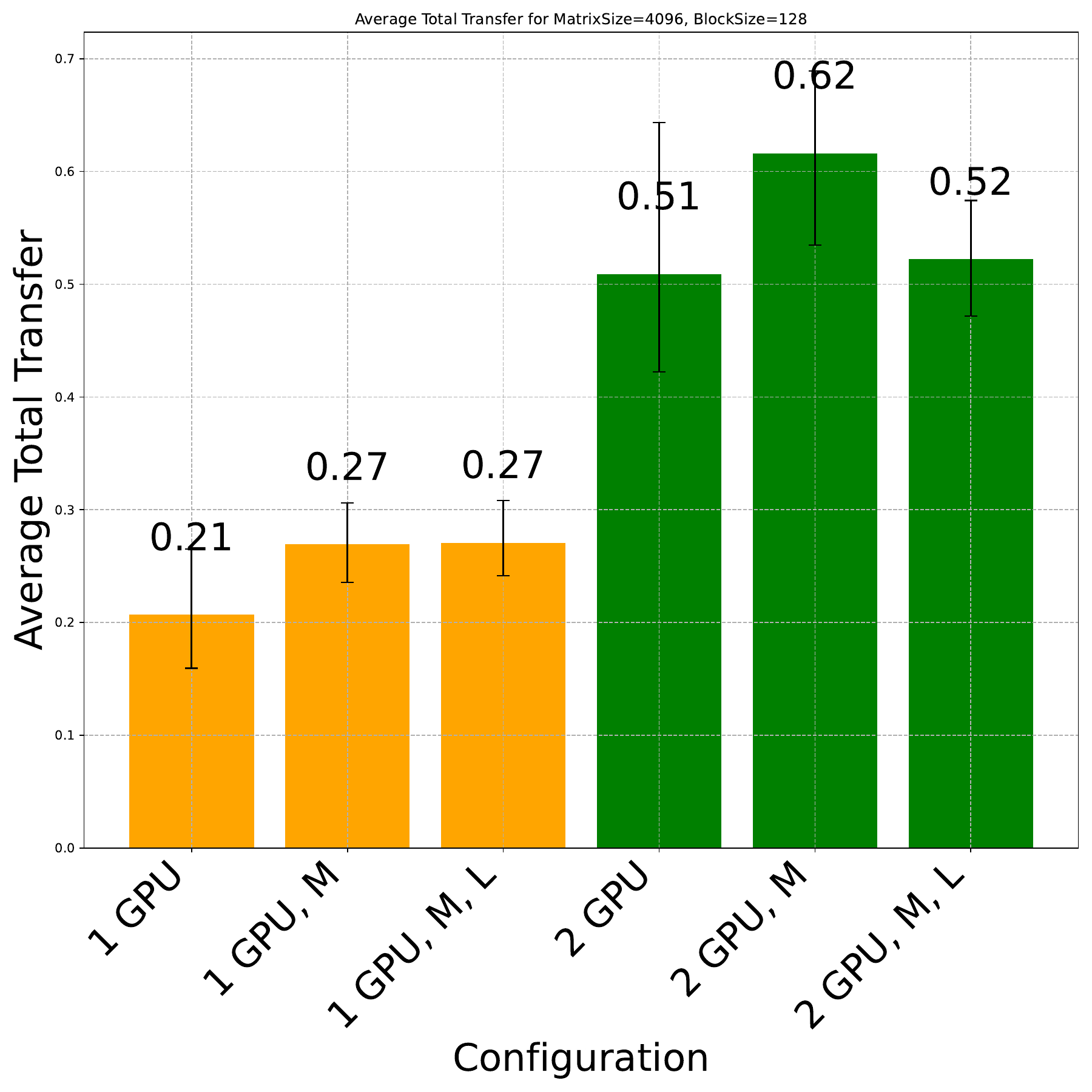}
        \caption{4096/128}
        \label{fig:transfers:4096_128}
    \end{subfigure}
    \hfill
    \begin{subfigure}{0.30\textwidth}
        \centering
        \includegraphics[width=\textwidth]{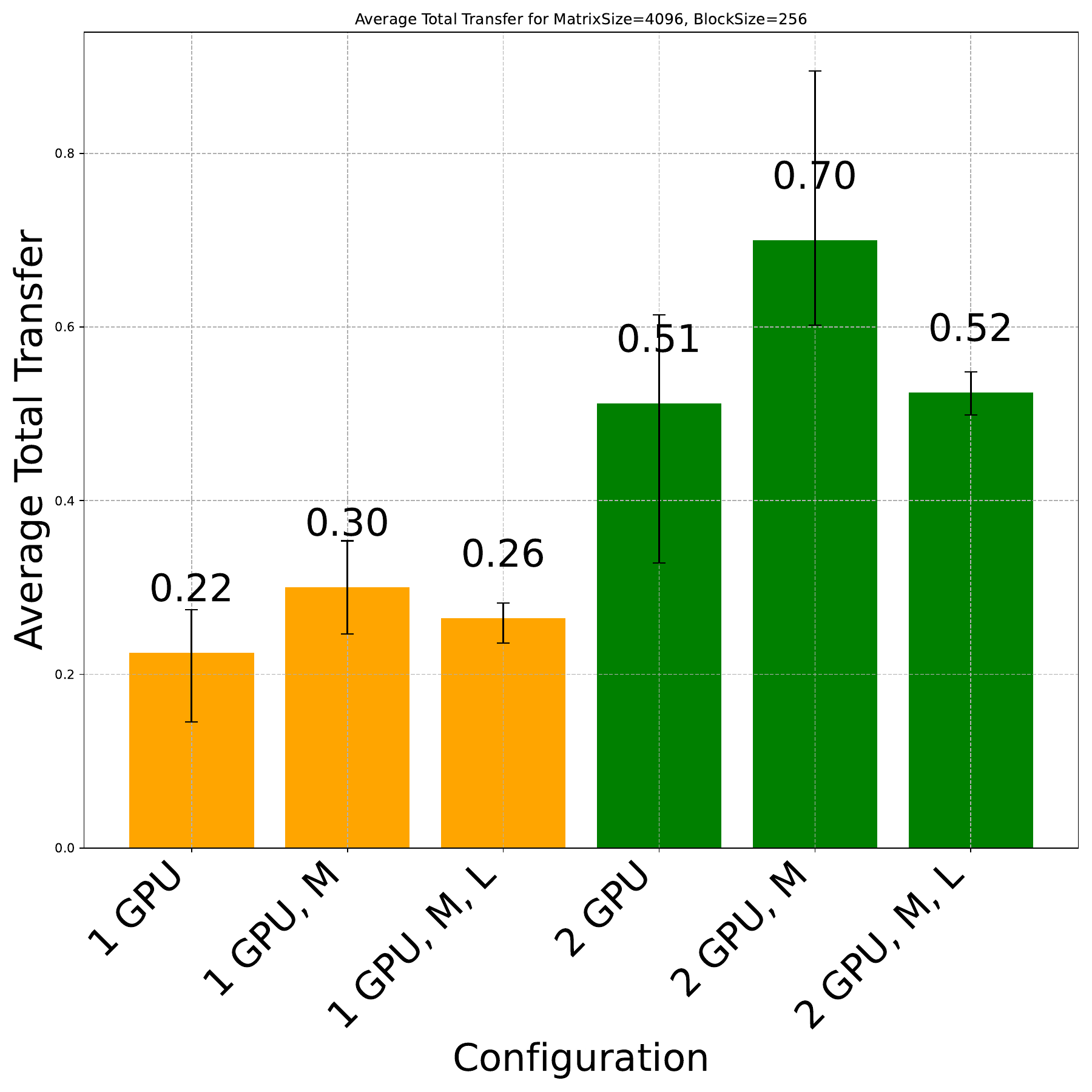}
        \caption{4096/256}
        \label{fig:transfers:4096_256}
    \end{subfigure}
    \hfill    
    \begin{subfigure}{0.30\textwidth}
        \centering
        \includegraphics[width=\textwidth]{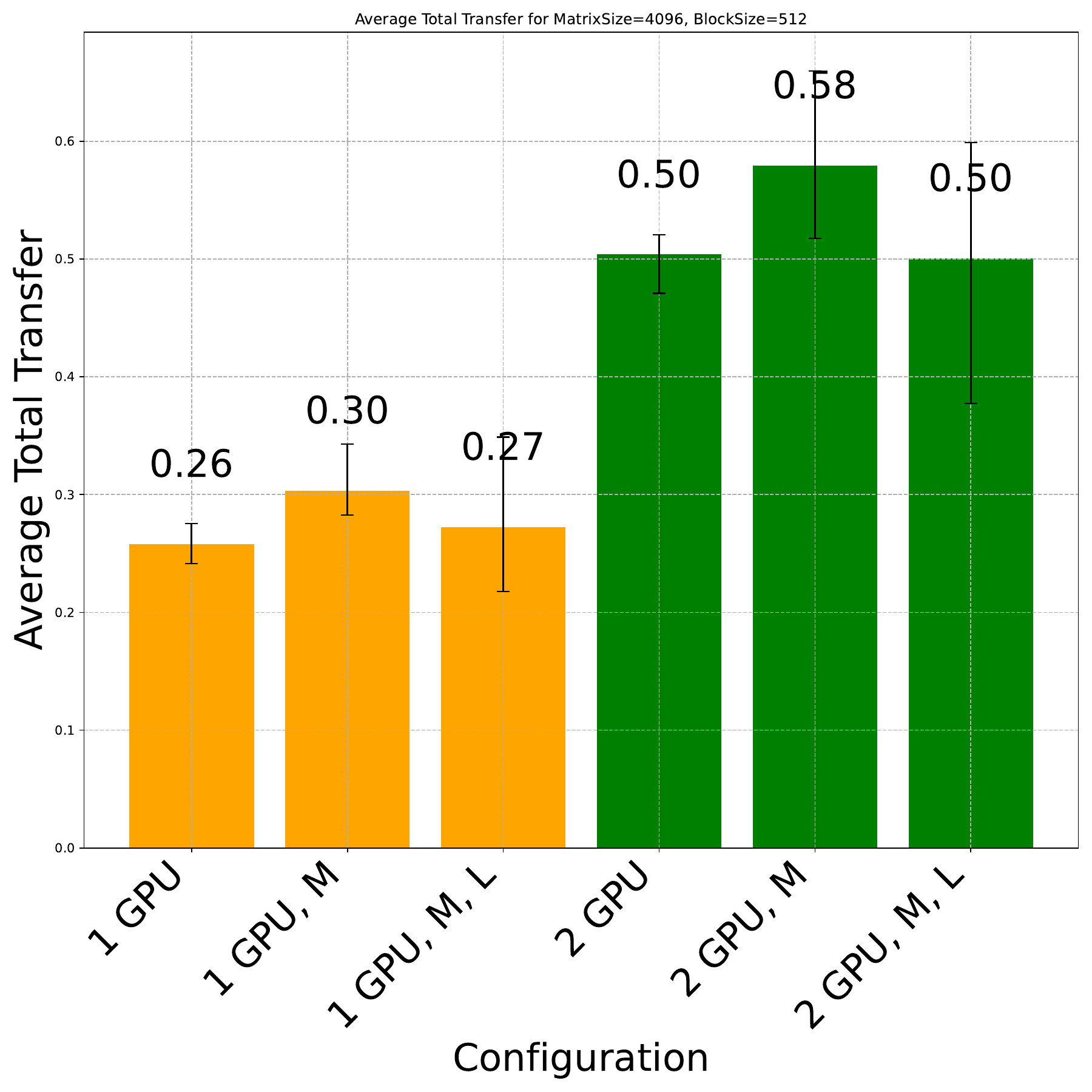}
        \caption{4096/512}
        \label{fig:transfers:4096_512}
    \end{subfigure}
    
    \begin{subfigure}{0.30\textwidth}
        \centering
        \includegraphics[width=\textwidth]{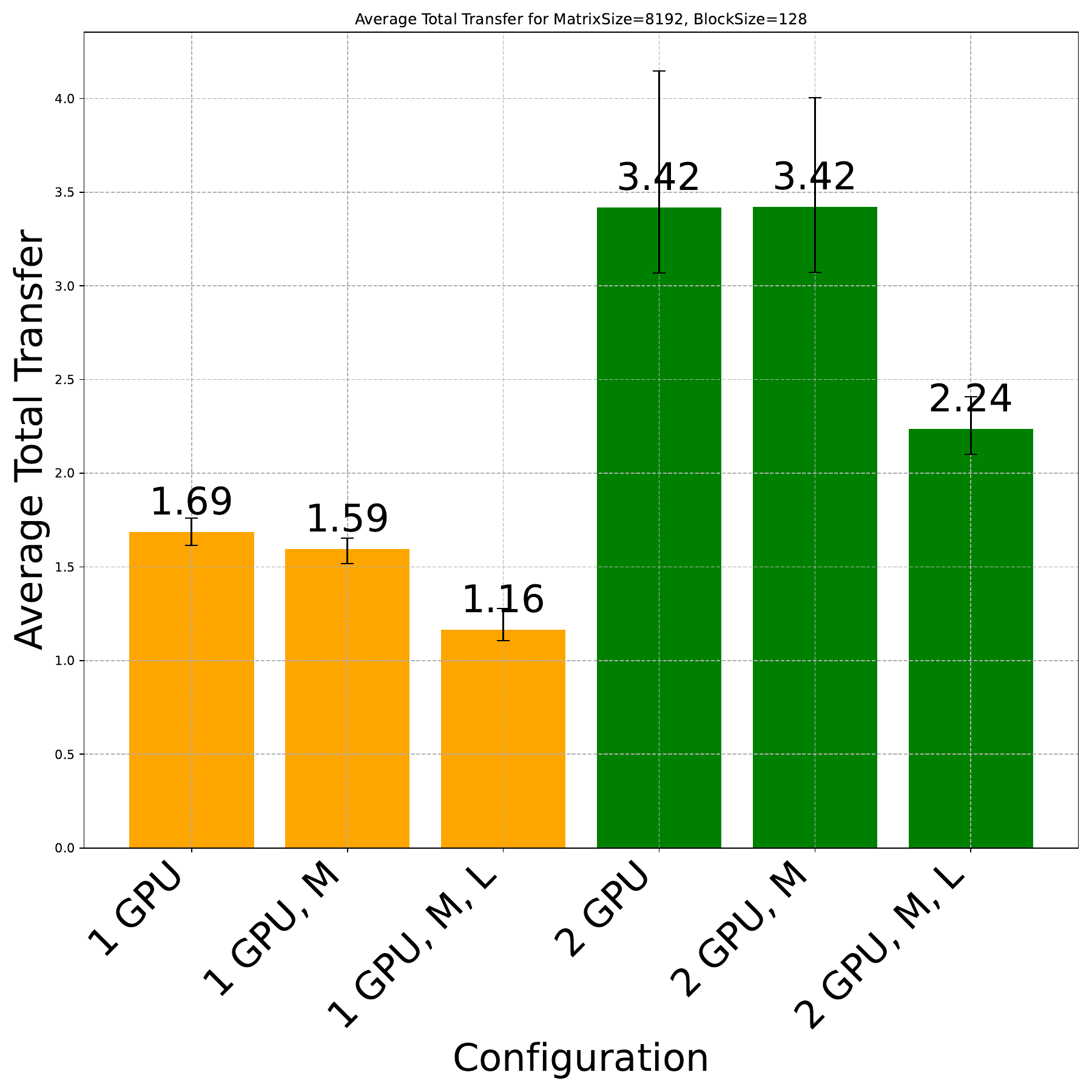}
        \caption{8192/128}
        \label{fig:transfers:8192_128}
    \end{subfigure}
    \hfill
    \begin{subfigure}{0.30\textwidth}
        \centering
        \includegraphics[width=\textwidth]{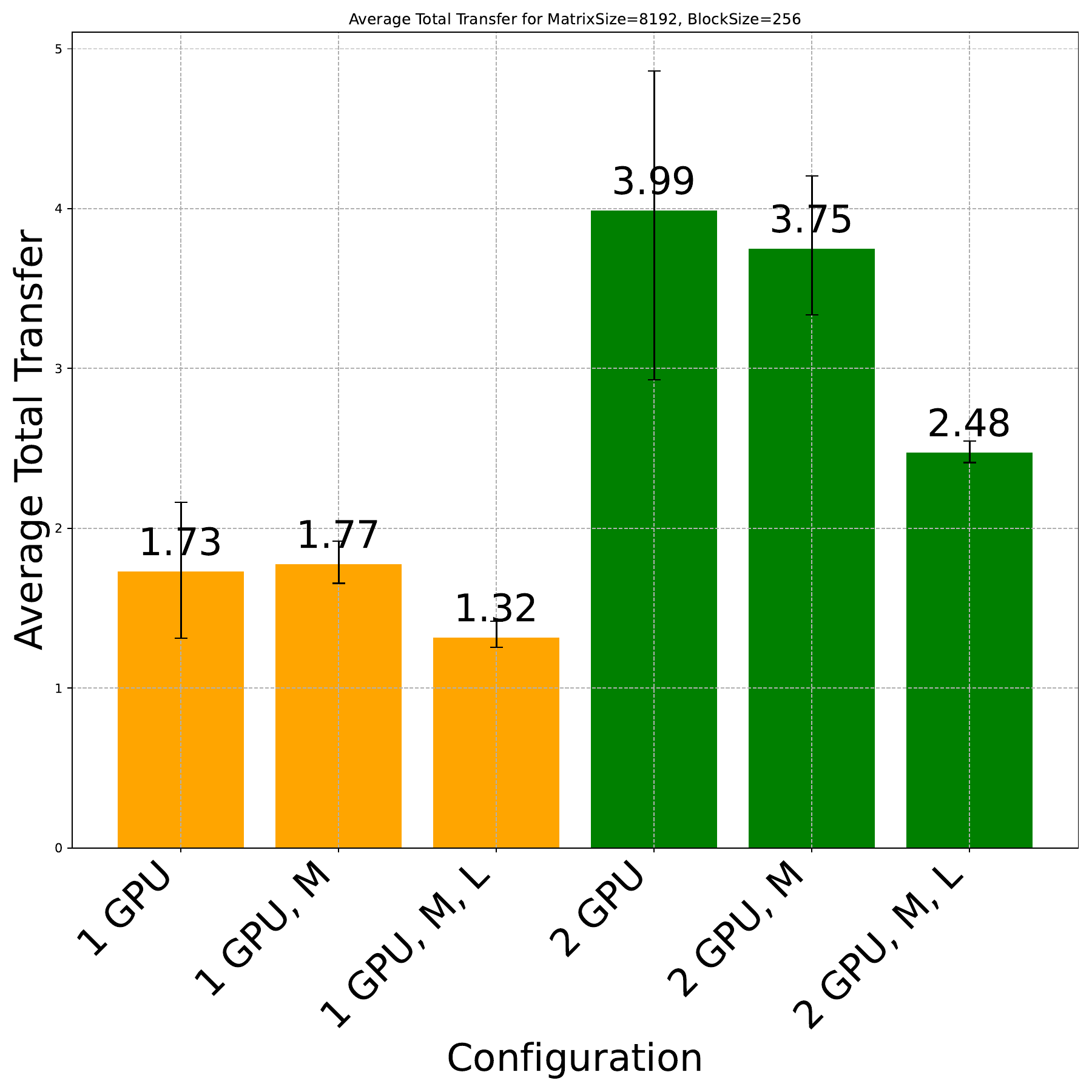}
        \caption{8192/256}
        \label{fig:transfers:8192_256}
    \end{subfigure}
    \hfill    
    \begin{subfigure}{0.30\textwidth}
        \centering
        \includegraphics[width=\textwidth]{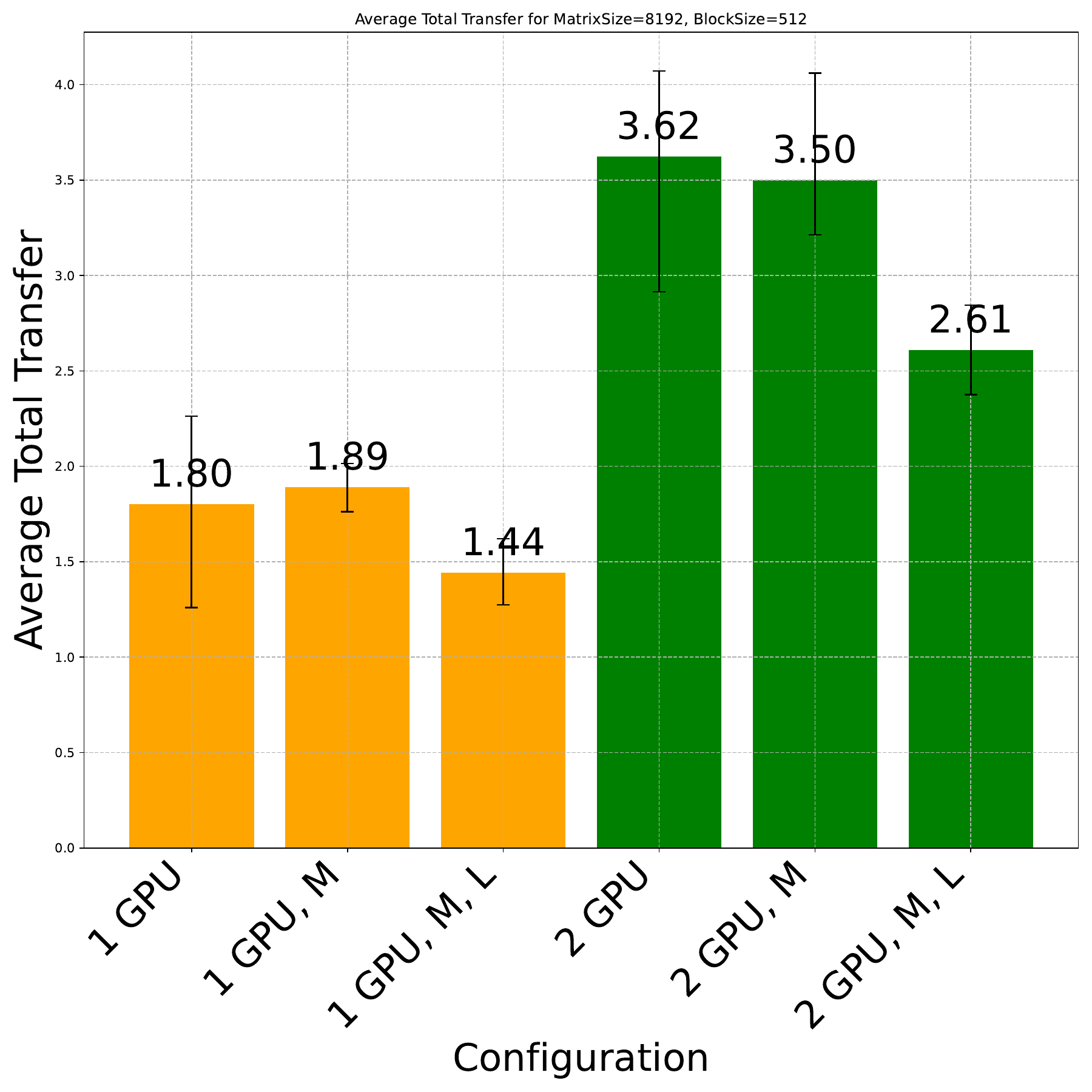}
        \caption{8192/512}
        \label{fig:transfers:8192_512}
    \end{subfigure}
    
    \begin{subfigure}{0.30\textwidth}
        \centering
        \includegraphics[width=\textwidth]{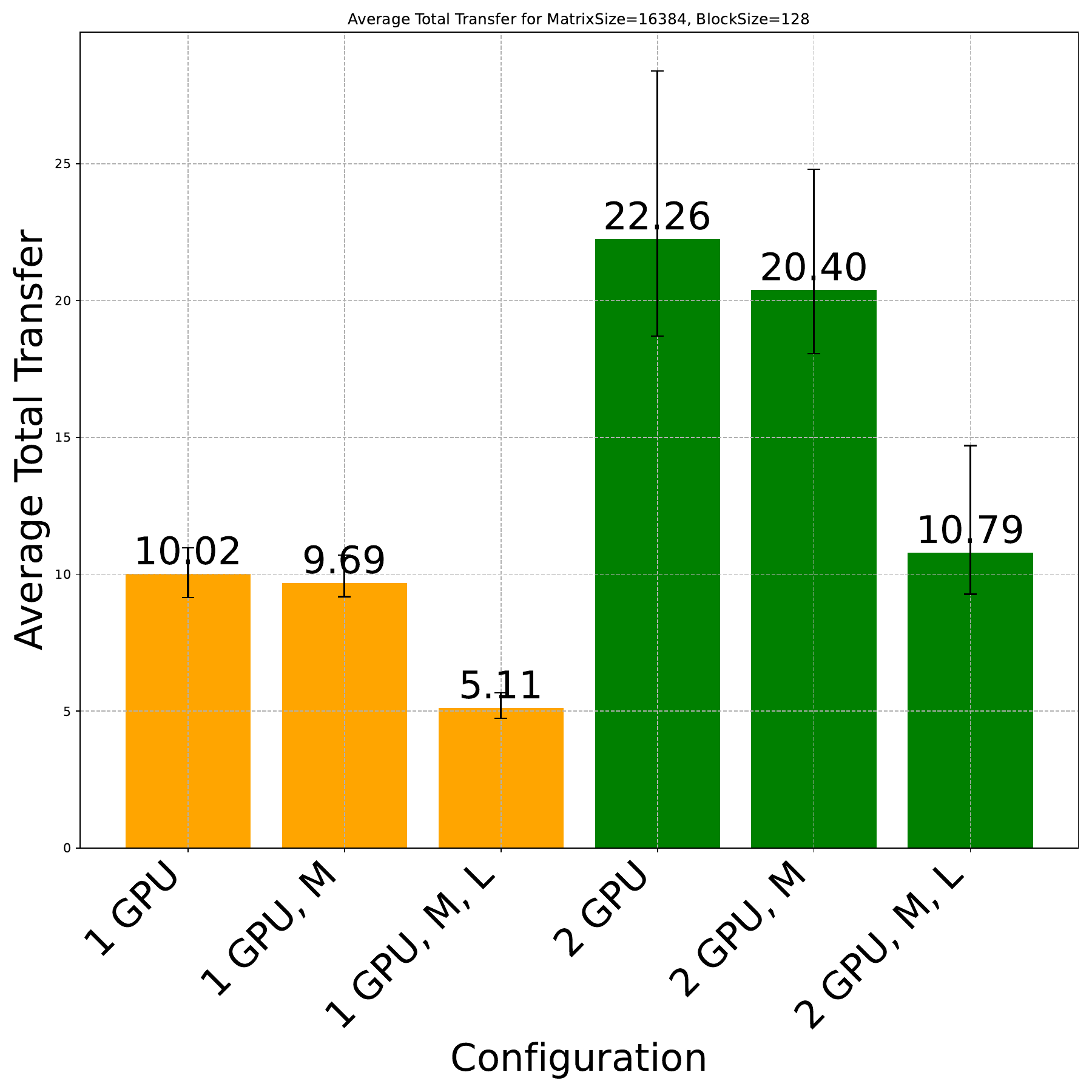}
        \caption{16384/128}
        \label{fig:transfers:16384_128}
    \end{subfigure}
    \hfill
    \begin{subfigure}{0.30\textwidth}
        \centering
        \includegraphics[width=\textwidth]{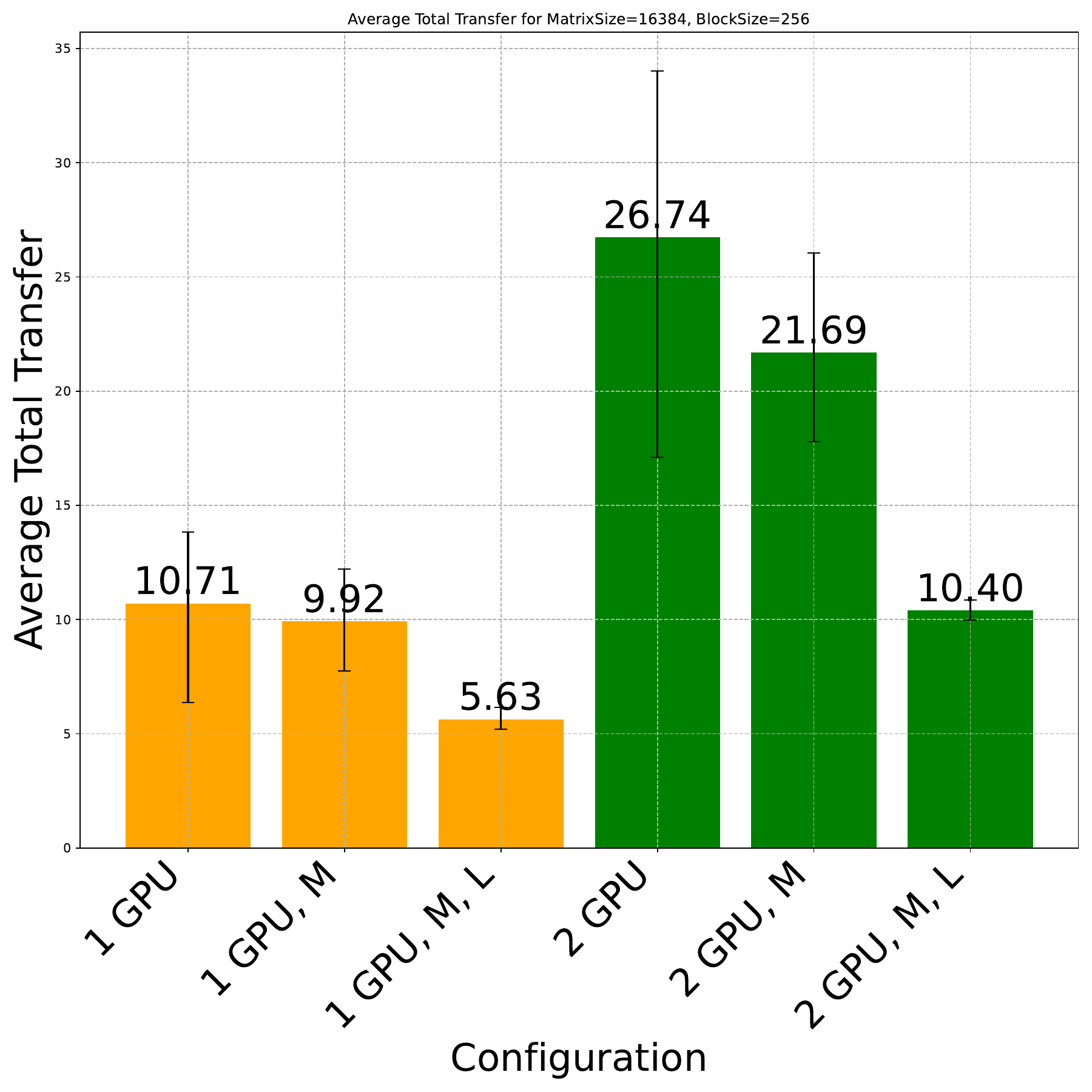}
        \caption{16384/256}
        \label{fig:transfers:16384_256}
    \end{subfigure}
    \hfill    
    \begin{subfigure}{0.30\textwidth}
        \centering
        \includegraphics[width=\textwidth]{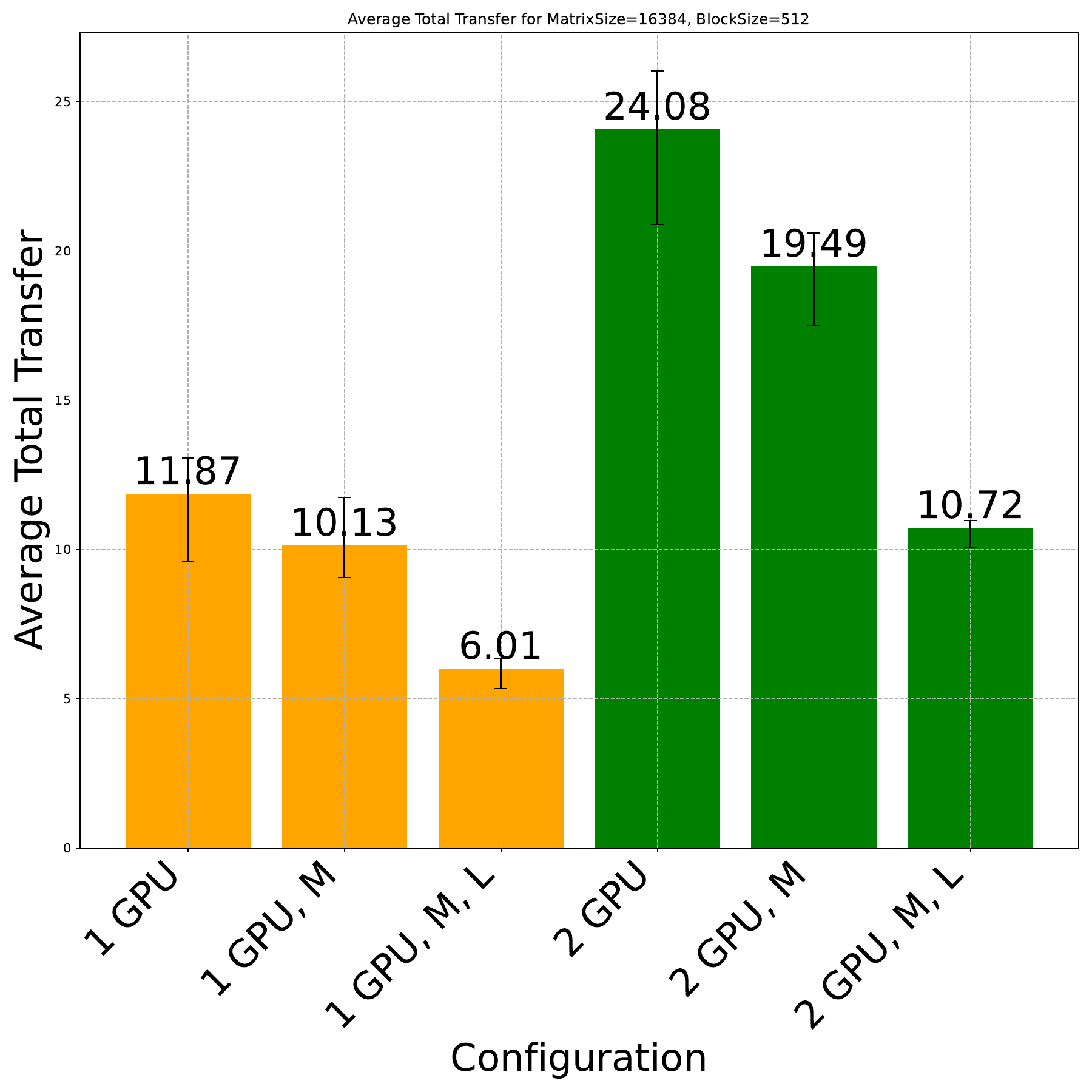}
        \caption{16384/512}
        \label{fig:transfers:16384_512}
    \end{subfigure}
    
    \caption{Transfers for the GEMM test case with different matrix sizes and block sizes,
            and different schedulers ($M$ for multriprio, $L$ for locality feature).}
    \label{fig:transfers}
\end{figure}

In Figure~\ref{fig:transfers}, we present the amount of memory transfers for the GEMM test case across various matrix sizes, block sizes, and schedulers. 
The data show that the number of transfers consistently increases when using two GPUs compared to one GPU, and the volume of transfers grows with matrix size, as expected.

Additionally, the number of transfers slightly increases with block size. 
This is because larger block sizes make GPUs relatively more efficient than CPUs. 
As a result, GPUs perform more computations and require more data to be transferred. 
Conversely, for small matrices, the performance benefit of GPUs is less pronounced; they compute fewer tasks, and the advantages of using the multriprio scheduler or the locality feature are diminished.

For large matrices, the multriprio scheduler generally provides better performance, and the locality feature is consistently beneficial. 
This improvement is due to the multriprio scheduler with locality, which ensures that each worker computes on the same $C$-panel.
Although this approach is not fully optimal, it reduces the number of transfers and enables potential reuse of $A$ or $B$-panel data.

All-in-all, the results show that the performance of Specx is highly dependent on the application and the hardware configuration.
The runtime system is capable of providing significant speedups, particularly when using GPUs, but the performance can be highly variable.
The scheduling choices can have a significant impact on the performance, and the data management can also be crucial.

\section{Conclusion}

We presented Specx, a task-based runtime system written in C++ and for C++ applications.
Specx allows parallelizing over distributed computing nodes and exploit CPUs and GPUs jointly.
It is easy to use and provides advanced features such as scheduler customization and execution trace visualization.
Performance study shows that Specx can be used for high-performance applications.

We plan to improve Specx by providing a scheduler made for heterogeneous computing node, create new speculative execution model, add conditional tasks, and improve the compilation error handling.
In addition, we would like to evaluate how C++ coroutines can be used to improve the performance or the design of the runtime system.

\section{Acknowledgement}
We used the PlaFRIM experimental testbed, supported by Inria, CNRS (LABRI and IMB), Universite de Bordeaux, Bordeaux INP and Conseil Regional d’Aquitaine~\footnote{\url{https://www.plafrim.fr}}. 

This work has been funded by the Inria ADT project SPETABARU-H, and the ANR National project AUTOSPEC (ANR-21-CE25-0009).

\bibliographystyle{plain} 

\end{document}